%% file: main.tex
\documentclass[journal]{IEEEtran}

\IEEEoverridecommandlockouts

\usepackage{cite}
\usepackage{pdfpages}
\usepackage[utf8]{inputenc}
\usepackage{amsmath,amssymb,amsfonts, mathrsfs}
\usepackage{mathabx}
\usepackage{dsfont}      
\usepackage{amssymb}
\usepackage{ragged2e}
\usepackage{tikz}
\usetikzlibrary{bayesnet}
\usepackage{algorithm}
\usepackage{algpseudocode}

\usepackage[normalem]{ulem}

\usepackage{euscript}

\usepackage{enumitem}
\usepackage{graphicx}
\usepackage{comment}
\usepackage{textcomp}
\usepackage{xcolor}
\usepackage{soul}
\usepackage{xfrac}
\usepackage{mathtools}
\usepackage{comment}
\usepackage{stmaryrd}
\usepackage{float}
\usepackage{booktabs}
\usepackage{tabularx}
\usepackage{balance}
\def\BibTeX{{\rm B\kern-.05em{\sc i\kern-.025em b}\kern-.08em
    T\kern-.1667em\lower.7ex\hbox{E}\kern-.125emX}}

\makeatletter
\newcommand*{\rom}[1]{\expandafter\@slowromancap\romannumeral #1@}
\usepackage{cuted} 
\usepackage{placeins}

\newcommand*{\hermconj}{^{\mathsf{H}}}
\newcommand*{\trans}{^{\mathsf{T}}}
\newcommand{\norm}[1]{\big\lVert#1\big\rVert}
\newcommand{\normF}[1]{{\big\lVert#1\big\rVert}_\mathsf{F}}

\usepackage[acronym,shortcuts]{glossaries}

\newcommand{\mycomment}[1]{}
\newcommand{\floor}[1]{\left\lfloor #1 \right\rfloor}
\usepackage{pgfplots}
\pgfplotsset{compat=newest}
\pgfplotsset{plot coordinates/math parser=false}
\newlength\fheight
\newlength\fwidth
\usetikzlibrary{plotmarks,patterns,decorations.pathreplacing,backgrounds,calc,arrows,arrows.meta,spy,matrix}
\usepgfplotslibrary{patchplots,groupplots}
\usepackage{tikzscale}

\DeclareMathOperator{\Tr}{\mathsf{tr}}
\DeclareMathOperator{\diag}{\mathsf{diag}}
\DeclareMathOperator{\vect}{\mathsf{vec}}
\DeclareMathOperator*{\argmax}{arg\,max}
\DeclareMathOperator*{\argmin}{arg\,min}

\newacronym{als}{ALS}{alternating least squares}
\newacronym{amp}{AMP}{approximate message passing}
\newacronym{ard}{ARD}{automatic rank determination}
\newacronym{cpd}{CPD}{canonical polyadic decomposition}
\newacronym{crc}{CRC}{cyclic redundancy check}
\newacronym{dvb}{DVB}{discrete variational Bayesian}
\newacronym{elbo}{ELBO}{evidence lower bound}
\newacronym{fasura}{FASURA}{fading spread unsourced random access}
\newacronym{fec}{FEC}{forward error correction}
\newacronym{gmac}{GMAC}{Gaussian multiple access channel}
\newacronym{kl}{KL}{Kullback-Leibler}
\newacronym{llr}{LLR}{log-likelihood ratio}
\newacronym{mgp}{MGP}{multiplicative Gamma process}
\newacronym{mimo}{MIMO}{multiple input-multiple output}
\newacronym{ml}{ML}{maximum likelihood}
\newacronym{mmtc}{mMTC}{massive machine-type communications}
\newacronym{mrc}{MRC}{maximum ratio combining}
\newacronym{noma}{NOMA}{non-orthogonal multiple access}
\newacronym{odma}{ODMA}{on-off division multiple access}
\newacronym{ptura}{PTURA}{polar-coded tensor URA}
\newacronym{pupe}{PUPE}{per-user probability of error}
\newacronym{sic}{SIC}{successive interference cancellation}
\newacronym{snr}{SNR}{signal-to-noise ratio}
\newacronym{tbm}{TBM}{tensor-based modulation}
\newacronym{ura}{URA}{unsourced random access}
\newacronym{dvbals}{DVB-ALS}{}
\newacronym{dvbtbm}{DVB-TBM}{}

\title{\huge{Variational Bayesian Tensor Decomposition With\\Discrete Mixture Prior for Unsourced Random Access}}

\author{Ala~Baccar\IEEEauthorrefmark{1}\IEEEauthorrefmark{2},
        Alexis~Decurninge\IEEEauthorrefmark{1},
        Alberto~Rech\IEEEauthorrefmark{1},
        Sofiane~Kharbech\IEEEauthorrefmark{1},
        Éric~Pierre~Simon\IEEEauthorrefmark{3},
        and~Joumana~Farah\IEEEauthorrefmark{2}\\[3mm]
\small
\IEEEauthorrefmark{1} Advanced Wireless Technologies Lab, Fourier Research Center, Huawei Technologies, Paris, France.\\
\IEEEauthorrefmark{2} University of Rennes, INSA Rennes, CNRS, IETR UMR 6164, Rennes, France.\\
\IEEEauthorrefmark{3} Université de Lille, CNRS, IEMN UMR 8520, Lille, France. \\
Corresponding author: A.~Baccar (email: ala.baccar@huawei.com).
}

\begin{document}
\maketitle

\IEEEoverridecommandlockouts
\bstctlcite{IEEEexample:BSTcontrol}

\begin{abstract}
\Ac{tbm} schemes are a promising approach for \ac{ura}, where user separation relies on decomposing the received signal tensor via the \ac{cpd}, typically computed with \ac{als}. Standard \ac{als}, however, treats the factor matrices as unstructured and fails to exploit the discrete structure of the tensor sub-constellations. 
We propose \ac{dvbals}, a discrete variational Bayesian \ac{cpd} framework with specific priors, tailored to a tensor structure with the corresponding encoding strategy, combined with iterative computation of an approximate posterior distribution. A discrete Gaussian mixture prior on one Grassmannian factor softly aligns the estimates toward the constellation points. The remaining factors are jointly modeled with a structured Gaussian prior whose posterior mean is constrained to the Khatri-Rao product manifold and posterior variance upper-bounded to prevent norm divergence during inference. The resulting closed-form coordinate ascent algorithm jointly estimates all latent factors and their uncertainties.
We integrate \ac{dvbals} into \ac{dvbtbm} to design a complete \ac{ura} receiver with single-user demapping, polar decoding with \ac{crc} verification, and \ac{sic}. Simulation results show significant gains over standard \ac{als}-based decomposition and robust detection performance in \ac{ura} settings, outperforming state-of-the-art schemes under high system loads.
\end{abstract}

\begin{picture}(0,0)(0,-440)
\put(0,0){
\put(0,0){\qquad \qquad \quad This paper has been submitted to IEEE for publication. Copyright may change without notice.}}
\end{picture}


\glsresetall

\section{Introduction}

The rapid evolution of wireless communication systems has led to increasingly diverse service requirements. Beyond traditional broadband services, emerging applications such as \ac{mmtc} and large-scale sensing networks~\cite{Wu2020Massive} require access mechanisms that can support a very large number of devices that are only sporadically active and typically transmit short data packets. In such scenarios, conventional grant-based access schemes become inefficient because of high signaling overhead and the need for prior user identification. Although grant-free access schemes~\cite{Liu2018Sparse} allow devices to transmit both metadata and data directly without requiring a prior resource request, they still face significant challenges. In particular, their performance is fundamentally limited by the need for orthogonal pilot assignment~\cite{Chen2017Capacity}.

\Ac{ura}~\cite{Polyanskiy2017perspective, Liva2024Unsourced} has emerged as a promising grant-free multiple access paradigm that addresses these limitations by shifting the receiver’s objective from identifying users to decoding transmitted messages. In \ac{ura}, devices send short messages selected from a common codebook without explicitly including their identities. The receiver aims to recover the set of transmitted messages, regardless of which devices sent them. This formulation enables scalable access for massive device populations and introduces new challenges for coding, decoding, and resource allocation design. Consequently, \ac{ura} has attracted considerable research interest as a key enabler for sporadic, low-latency, and grant-free communications in next-generation wireless systems~\cite{Chen2021Massive}.

Early \ac{ura} schemes for the quasi-static fading regime exploited the sparsity of the received signal to recover the users' messages~\cite{Fengler2021SPARCs, Fengler21Pilot}, typically relying on approximate message passing algorithms to solve the resulting compressed sensing problem at the \ac{mimo} receiver. 
One prominent scheme is \ac{fasura}~\cite{Gkagkos2023FASURA}, which sets the benchmark for energy efficiency in the \ac{ura} literature. It splits the resources into a short preamble, used for user activity detection and channel estimation via an energy detector, and a payload encoded and transmitted with a \ac{noma} scheme and recovered by coherent detection.
Building on this line of work, a related \ac{odma}~\cite{Ozates2024ODMA} scheme employs iterative orthogonal matching pursuit for activity detection and an on-off division \ac{noma} technique for the coherent part.

Beyond compressed sensing, tensor models have been widely adopted across wireless communication applications~\cite{Rech2026Tensor, Han2021Sparse, Kang2026Tensor}. 
Within the \ac{ura} framework, \ac{tbm} schemes have received considerable attention~\cite{Decurninge2021Tensor, Rech2023Unsourced, Baccar2025Tensor, Fang2025Polar, maxime2026Belief}: encoded sequences are distributed across the modes of a tensor, each mode drawn from a discrete sub-constellation, and the factors are combined via the Kronecker product to yield a received signal expressed as a sum of rank-one tensors. At the receiver, a \ac{cpd}-based algorithm~\cite{Kolda2009Tensor} separates the contributions of different users, followed by single-user demapping and \ac{sic}.

An alternative to conventional optimization-based tensor decomposition methods~\cite{Sidiropoulos2017Tensor, Kolda2009Tensor, cohen2018Dictionary} is the Bayesian approach, which infers posterior distributions of the model parameters rather than point estimates of the factor matrices. Bayesian methods~\cite{Tzikas2008variational} provide not only parameter estimates but also a measure of their uncertainty, enabling a more reliable assessment of factor identifiability in noisy and underdetermined settings. A widely studied instance is jointly promoting sparsity and rank determination~\cite{Zhao2015Bayesian}, which uses classical Gaussian-Gamma priors to model sparsity and infer the tensor rank.
The choice of prior is central to the performance of such frameworks, and several works refine it toward more accurate rank determination. In~\cite{Hiroki2022Bayesian}, the conventional Gamma prior on the precision parameters is replaced by a multiplicative Gamma process, reducing redundancy in the factor columns and sharpening the rank estimate. Similarly, \cite{Cheng2022Towards} investigates hyperbolic priors on the factor matrices as an alternative means of promoting sparsity. Beyond rank determination, \cite{Sun2024Bayesian} applies Bayesian inference based on the alternate prior hypothesis to channel estimation in \ac{mimo}-OFDM systems, explicitly modeling distinct interference types to jointly estimate the number of propagation paths together with the channel and interference statistics.

These Bayesian principles have recently been extended to \ac{tbm} schemes for \ac{ura}, integrating polar coding with Bayesian inference. In~\cite{Fang2025Polar}, the \ac{ptura} scheme is proposed, wherein Gaussian priors on the factor matrices promote sparsity, enabling \ac{ard} and improving soft-decoding performance, with mean-field variational inference used to estimate the posteriors of the latent variables. \Ac{sic} is incorporated within the Bayesian framework by partitioning the observations into recovered and unrecovered components. These enhancements yield significant gains over conventional \ac{als}.
The tensor decomposition algorithm is thus a key component of \ac{tbm} schemes for \ac{ura}. Yet, most existing approaches, including~\cite{Fang2025Polar}, do not account for the discrete nature of the sub-constellations to which the tensor factors belong. To address this, some of the authors of this paper proposed a hybrid tensor decomposition algorithm~\cite{Baccar2025Tensor}, in which the last factor embeds a small sub-constellation, and this property is exploited by hard-projecting the continuous estimates onto the corresponding sub-constellation via a \ac{ml} criterion during the iterations of the \ac{als} procedure. This discrete constraint improves the handling of large user loads. Still, the resulting scheme is suboptimal, unstable at low \ac{snr}, and sensitive to hyperparameter tuning used to evaluate the quality of the projected factor. It does not provide natural uncertainty information on the estimated parameters.

In this paper, we replace the hard projection with a principled probabilistic mechanism through a \ac{dvb} approach. A Gaussian mixture prior centered at the constellation points yields a soft attraction toward the sub-constellation while preserving uncertainty, improving robustness at low \ac{snr} and under heavy multi-user interference. Similar priors have been used in other contexts, such as sparse signal recovery~\cite{Jisheng2023Sparse} and block term decomposition~\cite{kofidis2024Revisiting}.
The main contributions of this work, centered on the variational Bayesian framework \ac{dvbals}, are summarized as follows.
\begin{itemize}[leftmargin=*, itemsep=4pt, topsep=4pt]
\item \textbf{Discrete-structure Bayesian factor posterior.} We propose a variational Bayesian \ac{cpd} in which a discrete mixture prior is placed on one factor matrix, probabilistically enforcing alignment with the predefined Grassmannian sub-constellation. This softly attracts the estimated factors toward the constellation points during the updates and, unlike hard projection, preserves posterior uncertainty and improves robustness to noise and multi-user interference.
\item \textbf{Structured Gaussian variational posterior.} To avoid the normalization problem inherent to the tensor structure \cite{Fang2025Polar}, the variational posterior over the latent tensor factors, excluding the discrete factor, is constrained to be a common Gaussian distribution with constraints on the mean and variance: the mean is structured as a Khatri-Rao product of the factor matrices, and the expected norm of the posterior is upper-bounded by a  \textit{signal energy} value to prevent collapse and norm explosion.
\item \textbf{\ac{dvbals} at the \ac{ura} receiver (\ac{dvbtbm}).} We integrate \ac{dvbals} into a complete \ac{ura} receiver that performs user
separation via Bayesian tensor factorization, followed by single-user demapping, polar decoding with \ac{crc} verification, and \ac{sic}. We call the proposed scheme \ac{dvbtbm}.
\end{itemize}

The rest of the paper is organized as follows. In Section~\ref{sec:system_model}, we introduce the system model, the tensor-based encoding scheme, and the optimization problem at the receiver. Section~\ref{sec:bayesian} discusses the Bayesian framework and introduces the \ac{dvbals} algorithm. The full receiver \ac{dvbtbm} architecture is presented in Section~\ref{sec:dvbtbm}. Performance evaluation of the proposed scheme is assessed in Section~\ref{sec:num_res}. Finally, Section~\ref{sec:conclusions} draws the main conclusions.

\emph{Notation. }%
$x$ or $X$, $\mathbf{x}$, $\mathbf{X}$, $\mathcal{X}$, $\EuScript{X}$ are respectively used for scalars, vectors (column vectors), matrices, $n$-dimensional arrays ($n>2$), and sets $\centerdot$
$\mathbf{I}_n$ identity matrix of size $n$ $\centerdot$
$\otimes$ Kronecker product $\centerdot$
$\odot$ Khatri–Rao product $\centerdot$
$\llbracket . \rrbracket$ tensor yielded from the factor matrices in argument $\centerdot$
$\big|\{.\}\big|$ cardinality of the set $\{.\}$  $\centerdot$
$\mathbb{E}[.]$ expectation operator $\centerdot$
$(.)\hermconj$ Hermitian transpose $\centerdot$
$(.)^*$ complex conjugate $\centerdot$
$(.)\trans$ transpose $\centerdot$
$(.)^{-1}$  matrix inverse $\centerdot$
$\vect(.)$ vectorization $\centerdot$
$\norm{.}$ $\ell^2$-norm;
$\normF{.}$ the Frobenius norm $\centerdot$
$\diag(\cdot)$ a diagonal matrix formed from its arguments $\centerdot$
$\Tr(\mathbf{A})$ denotes the trace of $\mathbf{A}$ $\centerdot$
$\mathbb{I}(\cdot)$ denotes the indicator function $\centerdot$ $\mathbf{x} \sim \mathcal{CN}(\boldsymbol{\mu},\mathbf{\Omega})$ denotes the circularly symmetric complex Gaussian random vector with mean $\boldsymbol{\mu}$ and covariance matrix $\mathbf{\Omega}$ $\centerdot$
$\mathcal{G}(x;a,b)$ Gamma distribution with shape $a$ and rate $b$  $\centerdot$
$\floor{.}$ is the floor function $\centerdot$
$\underline{\mathrm{c}}$ denotes an arbitrary constant term.

\section{System Model}
\label{sec:system_model}

We consider an uplink transmission system consisting of a large number $K$ of single-antenna users and one receiver with $N$ antennas. A subset of users $\EuScript{K}_\mathrm{a}$ is randomly activated and accesses the receiver in an uncoordinated manner. 
Without loss of generality, we assume $\EuScript{K}_\mathrm{a} = \{1, \dots, K_{\rm a}\}$ where $K_\mathrm{a}$ is the number of active users, which is assumed to be known at the receiver.
Each active user $k$ generates a binary message $\mathbf{m}_k$ which is encoded and modulated into a signal $\mathbf{x}_k \in \mathbb{C}^T$, normalized such that $\|\mathbf{x}_k\|^2 = T$. It is transmitted over the same set of $T$ available resources through a quasi-static Rayleigh fading channel $\mathbf{h}_k \in \mathbb{C}^N$, i.e., each entry of $\mathbf{h}_k$ is independently drawn from a standard Gaussian distribution.
The signal at the receiver is expressed as 
\begin{equation}
\mathbf{Y}=\sum_{k = 1}^{K_{\rm a}}\mathbf{x}_{k} \mathbf{h}_{k}\trans+\mathbf{Z},
\end{equation}
where $\mathbf{Z}\in \mathbb{C}^{T\times N}$ is the white complex-valued Gaussian noise with independent and identically distributed (i.i.d.) entries, each drawn from a circularly symmetric complex Gaussian distribution $\mathcal{CN}\left(0, \sigma^2\right)$ with variance $\sigma^2$.

Upon the uplink transmissions, the receiver produces a list $\widehat{\EuScript{L}}=\{\widehat{\mathbf{m}}_k:\, k =1,\ldots, \widehat K_{\rm a}\}$ of $\widehat K_{\rm a}$ decoded messages, which ideally matches the list of transmitted packets $\EuScript{L}$. The main objective of the \ac{ura} scheme design is to maximize the message retrieval capability.

\subsection{Tensor-based Encoding Strategy}
\label{sec:encoder}

The message encoding procedure follows the unsourced Hybrid-\ac{tbm} scheme~\cite{Baccar2025Tensor}, which maps information bits into structured tensor codewords. Let $d$ denote the total number of tensor modes and $T_i$ the dimension of the $i$-th mode, such that $\prod_{i=1}^{d} T_i = T$. The binary message $\mathbf{m}_k$ of user $k$ is partitioned into two parts. The first part, intended for the first $d-1$ modes, is \ac{fec}-encoded and appended with \ac{crc} bits to form the sequence $\mathbf{m}_k^\prime$. This sequence is then divided into $d-1$ disjoint subsets and mapped onto the factors $\mathbf{a}_{k,i}\in\mathbb{C}^{T_i}$, $i=1,\ldots,d-1$, using the CubeSplit Grassmannian constellation~\cite{Ngo2020Cube}, where each factor is a codeword selected from the sub-constellation $\EuScript{C}_i\subset\mathbb{C}^{T_i}$ associated with the $i$-th tensor mode. Each factor is normalized to have norm $\sqrt{T_i}$. The second part consists of $z$ bits assigned to the last tensor mode, which bypass channel encoding. These bits are directly mapped onto the factor $\mathbf{c}_k\in\EuScript{C}_d$, whose codewords are normalized to have norm $\sqrt{T_d}$. At the receiver, this factor is recovered using \ac{ml} detection.

The $d$-th factor $\mathbf{c}_{k}$ is denoted separately because it is the mode subject to discrete optimization: it is mapped from a small number of bits $z$, which keeps its discrete optimization within the \ac{cpd} tractable~\cite{Baccar2025Tensor}. The approach extends to any mode that encodes few bits, where it attains optimal performance; this condition does not hold for the other modes in the application considered here. The transmitted signal is then the rank-one codeword (written in its vectorized version)
\begin{equation}\label{eq:xk}
\mathbf{x}_k= \mathbf{a}_{k,1} \otimes \cdots \otimes \mathbf{a}_{k,d-1} \otimes \mathbf{c}_{k}
\;\in\; \mathbb{C}^T.
\end{equation}

\subsection{Induced Receiver Decoding Problem}
\label{sec:receiver_problem}

The \ac{cpd}-based receiver follows directly from this encoding strategy. Each active user transmits a rank-one tensor formed by the Kronecker product of its factor vectors, so the superposition over active users yields a low-rank tensor at the receiver. Augmented by the channel $\mathbf{h}_k$ as an additional mode, the received signal tensor is
\begin{equation}
\mathcal{Y} = \sum_{k = 1}^{K_{\rm a}} \mathbf{a}_{k,1} \otimes \cdots \otimes \mathbf{a}_{k,d-1} \otimes \mathbf{c}_{k} \otimes \mathbf{h}_k + \mathcal{Z},
\label{eq:sysmod2}
\end{equation}
where $\mathcal{Z}$ is the tensor form of the additive noise $\mathbf{Z}$. Modes $1,\dots,d-1$ carry the Grassmannian factors, mode $d$ the discrete factor $\mathbf{c}_k$, and mode $d+1$ the channel.

Without accounting for the \ac{fec} constraints, the \ac{ml} receiver consists of jointly estimating the factor vectors $\{\widehat{\mathbf{a}}_{k,i}, i=1\cdots d-1\}$, the channel vectors $\{\widehat{\mathbf{h}}_k\}$, and the discrete codewords $\{\widehat{\mathbf{c}}_k\}$ by solving
\begin{equation}
	\argmin_{\substack{\{\mathbf{a}_{k,i}\in \EuScript{C}_i \forall i\} \\ \{\mathbf{h}_{k}\in \mathbb{C}^{N}\} \\ \{ \mathbf{c}_{k} \in \EuScript{C}_d\}}}
	\left\lVert
	\mathcal{Y}
	-
	\sum_{k=1}^{K_\mathrm{a}}
	\mathbf{a}_{k,1}\otimes\cdots\otimes\mathbf{a}_{k,d-1}\otimes\mathbf{c}_{k}\otimes\mathbf{h}_{k}
	\right\rVert^2,
\label{eq:hard_form}
\end{equation}
which seeks the rank-one decomposition that best matches the observation while satisfying the structural constraints on the factors. A standard strategy to approximately solve~\eqref{eq:hard_form} is to relax the constraints of the factors to belong to continuous complex spaces, i.e., optimizing over $\{\mathbf{a}_{k,i} \in \mathbb{C}^{T_i} \forall i\}$ and $\{\mathbf{c}_{k} \in \mathbb{C}^{T_d}\}$. However, even with this relaxation, the optimization is challenging to solve, as low-rank tensor decomposition is non-convex and NP-hard~\cite{Vavasis2010On,Hillar2013Most}. A widely adopted approach is the \ac{als} algorithm, which updates one factor at a time by least squares while keeping the others fixed~\cite{Kolda2009Tensor}. Although efficient, conventional \ac{als} ignores the discrete nature of the factors, making it prone to estimation errors and overfitting in noisy conditions. Hybrid-\ac{als}~\cite{Baccar2025Tensor} addresses this issue by enforcing a discrete constraint through a hard \ac{ml} projection on one mode,
\begin{equation}
    \widehat{\mathbf{c}}_k \gets \argmin_{\mathbf{c} \in \EuScript{C}_d} \norm{\widehat{\mathbf{c}}^{\mathrm{LS}}_k - \mathbf{c}}^2 \quad \forall k,
    \label{eq:ml}
\end{equation}
where $\widehat{\mathbf{c}}^{\mathrm{LS}}_k$ is the \ac{als} least-squares estimate. The algorithm updates the continuous modes by least squares, computes $\mathbf{c}_k$ by least squares followed by the projection~\eqref{eq:ml}, and feeds the discretized $\widehat{\mathbf{c}}_k$ back into the remaining updates, iterating to convergence. While this process markedly improves decomposition capacity, it is unstable at low-to-moderate \ac{snr}~\cite{Baccar2025Tensor,cohen2018Dictionary}, where hard decisions amplify estimation errors by discarding uncertainty and rendering the optimization landscape non-smooth.
These limitations motivate changing the \ac{ml} point of view and considering the Bayesian approach presented in this paper.

\section{Discrete Bayesian Variational\\Tensor Decomposition}
\label{sec:bayesian}

In this section, we develop the Bayesian framework underlying the proposed approach and present the \ac{dvbals} algorithm. We first define the probabilistic model by specifying the prior distributions on the system parameters, then derive the variational inference procedure used to approximate the intractable posterior distributions, and finally summarize the resulting \ac{dvbals} algorithm.

\subsection{Bayesian Probabilistic Model}

\emph{Signal Model.}
Rather than making a single hard decision about each unknown variable, Bayesian inference assigns a probability to each possible outcome \cite{bishop2011Pattern}. By maintaining posterior distributions, Bayesian soft inference explicitly captures uncertainty in the estimation process. Rather than committing to a single discrete choice at each iteration, the model can defer hard decisions until sufficient contextual information is accumulated. This principled handling of ambiguity leads to improved numerical stability and more reliable convergence.
Define $\mathbf{A}_i \triangleq \big[\mathbf{a}_{1,i}, \cdots, \mathbf{a}_{k,i}, \cdots, \mathbf{a}_{K_\mathrm{a},i}\big] \in \mathbb{C}^{T_i\times K_\mathrm{a}}$ ($i=1, \cdots, d-1$), $\mathbf{C} \triangleq \big[\mathbf{c}_1, \cdots, \mathbf{c}_k, \cdots, \mathbf{c}_{K_\mathrm{a}} \big] \in \mathbb{C}^{T_d \times K_\mathrm{a}}$, and $\mathbf{H} \triangleq \big[\mathbf{h}_1, \cdots, \mathbf{h}_k, \cdots, \mathbf{h}_{K_\mathrm{a}} \big] \in \mathbb{C}^{N \times K_\mathrm{a}}$. 
The likelihood of the mode-$d$ unfolding $\mathbf{Y}^{(d)}$, conditioned on the factor matrices $\mathbf{A}_1, \ldots, \mathbf{A}_{d-1}$, $\mathbf{H}$, $\mathbf{C}$, and $\tau$, can be deduced from~\eqref{eq:sysmod2} as
\begin{equation}
\begin{aligned}
p\big(\mathbf{Y}^{(d)} \,\big|\,
\mathbf{A}_1, \dots, \mathbf{A}_{d-1}, \mathbf{H}, \mathbf{C}, \tau \big)
= \mathcal{CN}\big( &
\vect(\mathbf{Y}^{(d)}); \\
\vect(\mathbf{C}\mathbf{K}\trans), \, \tau^{-1}\mathbf{I}_{TN} \big),
\end{aligned}
\label{eq:likli}
\end{equation}
where
\begin{equation} 
\mathbf{K} = \mathbf{H} \odot \mathbf{A}_{d-1} \odot \mathbf{A}_{d-2} \odot \dots \odot \mathbf{A}_1,
\label{eq:kr}
\end{equation}
and $\tau^{-1}$ denotes the noise precision, i.e.,  the inverse of the noise variance $\tau=1/\sigma^2$. 
Note that for the sake of notation simplicity, we will omit in this section the dependencies between the density functions $p$ and the random variable they characterize since this dependency is clear from the context.
To construct the Bayesian framework, we define priors over the latent variables. 

\emph{Prior on $\mathbf{C}$.}
We first consider a hierarchical prior on $\mathbf{C}$. Specifically, a discrete mixture prior is first imposed on the factor matrix $\mathbf{C}$ to encourage each user-specific vector $\mathbf{c}_k$ to align with one of the discrete constellation points $\mathbf{v}_m \in \EuScript{C}_d$. The prior is expressed as
\begin{equation}
    p(\mathbf{C} \mid \mathbf{G}, \{\gamma_k\}) = \prod_{k=1}^{K_\mathrm{a}} \prod_{m=1}^{M} \mathcal{CN}\big(\mathbf{c}_k;\, \mathbf{v}_m,\, \gamma_k^{-1}\mathbf{I}_{T_d}\big)^{g_k[m]},
\end{equation}
where $M = |\EuScript{C}_d|=2^z$ denotes the cardinality of the discrete sub-constellation $\EuScript{C}_d$. The density $\mathcal{CN}(\mathbf{c}_k;\, \mathbf{v}_m,\, \gamma_k^{-1}\mathbf{I})$ represents a circularly symmetric complex Gaussian distribution with mean vector $\mathbf{v}_m $ and covariance matrix $\gamma_k^{-1}\mathbf{I}_{T_d}$. Here, $\mathbf{v}_m$ corresponds to the $m$-th point in the sub-constellation set, serving as the localized mean for the $k$-th user's latent vector when assigned to that component. When $\gamma_k$ is large, the covariance $\gamma_k^{-1}\mathbf{I}_{T_d}$ becomes small, forcing $\mathbf{c}_k$ to concentrate tightly around the selected constellation point $\mathbf{v}_m$, effectively enforcing discrete behavior. On the other hand, if $\gamma_k$ is small, the covariance becomes large, allowing $\mathbf{c}_k$ to deviate from the constellation points, making the estimate more influenced by the likelihood than the prior. The elements $g_k[m] \in \{0,1\}$ serve as binary indicators of the event that the user $k$ has transmitted the codeword $\mathbf{v}_m$. Indeed, $g_k[m] = 1$ means that the $m$-th point in the sub-constellation $\EuScript{C}_d$ is responsible for generating the latent vector $\mathbf{c}_k$. 
These vectors satisfy the one-hot constraint $\sum_{m=1}^{M} g_k[m] = 1$ for all $k$.  
Collecting all user vectors column-wise yields the matrix
\begin{equation}
\mathbf{G} =
\begin{bmatrix}
\mathbf{g}_1 , \mathbf{g}_2 ,\cdots , \mathbf{g}_{K_\mathrm{a}}
\end{bmatrix}
\in \{0,1\}^{M \times K_\mathrm{a}}.
\end{equation}

To complete the hierarchical model, we assign a categorical prior to the responsibility vectors $\mathbf{g}_k$, expressed as:
\begin{equation}
    p(\mathbf{g}_k) = \prod_{m=1}^{M} \rho_m^{g_k[m]},
\end{equation}
where $\rho_m$ denotes the prior probability of the $m$-th constellation component. A typical choice is to consider that all codewords have the same weight, i.e., $\rho_m = 1/M$. Note that an alternative is to use a Dirichlet prior to allow adaptive mixture weighting \cite{bishop2011Pattern}. Furthermore, each user-specific precision parameter $\gamma_k$ is assigned an independent Gamma prior,
\begin{equation}
    p(\boldsymbol{\gamma})
    = \prod_{k=1}^{K_\mathrm{a}} \mathcal{G}\!\big(\gamma_k;\, a_{\gamma,k},\, b_{\gamma,k}\big),
\end{equation}
where the Gamma distribution is defined as
\begin{equation}
\mathcal{G}(x;\,a,b) = \frac{b^{a}}{\Gamma(a)} x^{a-1} e^{-b x},\qquad x > 0,
\end{equation}
and $a_{\gamma,k}$ and $b_{\gamma,k}$ denote distribution parameters.
The choice of the categorical prior for $\mathbf{g}_k$, combined with the Gamma prior for the precision parameter $\gamma_k$, promotes proximity between the column vectors of $\mathbf{C}$ and the discrete alphabet, while maintaining a tractable framework for Bayesian inference. 

\emph{Prior on $\tau$.}
Similarly to $\gamma_k$, the noise precision $\tau$ is also endowed with a Gamma prior with distribution parameters $a_0$ and $b_0$, i.e.,
\begin{equation}
    p(\tau)
    = \mathcal{G}\!\big(\tau;\, a_0,\, b_0\big).
\end{equation}

\emph{Prior on $\mathbf{K}$.}
Since our primary objective is to probabilistically model the discrete behavior of the Grassmannian tensor mode $\mathbf{C}$, the remaining factors are jointly modeled through their Khatri-Rao product $\mathbf{K}$~(\ref{eq:kr}), treated as a single latent variable to account for collective uncertainty when estimating $\mathbf{C}$ and $\tau$. We impose a joint Gaussian prior on its columns:
\begin{equation}
    p(\mathbf{K}) = \prod_{k=1}^{K_\mathrm{a}} \mathcal{CN}(\mathbf{k}_k ; \mathbf{0}, \kappa \mathbf{I}_{T_K}),
\label{eq:k_prior}
\end{equation}
 where $\mathbf{k}_k$ is the $k$-th column of $\mathbf{K}$, $T_K = N \prod_{i=1}^{d-1} T_i$, and $\kappa$ is a hyperparameter controlling the prior variance. Since $\mathbf{k}_k$ combines the effect of the channel and the symbol vectors $\mathbf{a}_{k,i}$, $\kappa$ captures the expected received power of their combined effect. Note that we chose a prior variance common to all users since all users face the same fading distribution. A simple relaxation of this choice is to consider a prior variance per user if we have access to more information regarding the received signal power per user. We will discuss the practical choice of $\kappa$ in Section~\ref{sec:num_res}.

\emph{Bayesian prior summary.}
As a result of the above formulation, the Bayesian formulation considers the set of latent variables $\boldsymbol{\Theta} = \{\mathbf{C},\,\mathbf{K},\mathbf{G},\, \boldsymbol{\gamma},\, \tau\}$. 
\begin{figure}[t]
    \centering
    \input{figs/graph_model}
    \caption{Graphical representation of the Bayesian framework.}
\label{fig:bay_fram}
\end{figure}
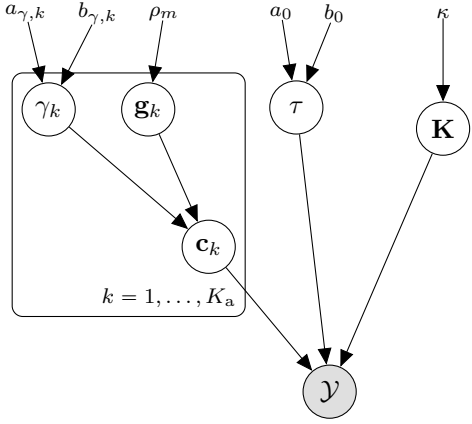
A graphical representation of the proposed hierarchical Bayesian model is illustrated in Fig.~\ref{fig:bay_fram}. 
The joint distribution of the observed tensor $\mathcal{Y}$ is 
\begin{equation}
\begin{aligned}
p(\mathcal{Y}, \boldsymbol{\Theta})
&=
p\big(\mathcal{Y} \mid \mathbf{C},\mathbf{K},\tau\big) \\
&\quad\times
p(\mathbf{C} \mid \mathbf{G}, \boldsymbol{\gamma})\,
p(\mathbf{G})\,
p(\mathbf{K})
p(\boldsymbol{\gamma})\,
p(\tau).
\end{aligned}
\label{eq:joint}
\end{equation}

\subsection{Bayesian Variational Inference}
Exact Bayesian inference requires evaluating the posterior distribution
\begin{equation}
p(\boldsymbol{\Theta} \mid \mathcal{Y}) 
= \frac{p(\mathcal{Y}, \boldsymbol{\Theta} )}{p(\mathcal{Y} )}.
\end{equation}
However, computing the marginal likelihood involves a combination of multidimensional integrals over $\mathbf{C}$, $\mathbf{K}$, $\boldsymbol{\gamma}$, and $\tau$, as well as summations over all possible configurations of the discrete assignment matrix $\mathbf{G}$. The dimensionality of these operations grows rapidly with the number of users and symbols, rendering exact Bayesian inference computationally intractable.
To address this intractability, variational mean-field inference is adopted~\cite{bishop2011Pattern}. The core idea is to approximate the true posterior distribution
$p(\boldsymbol{\Theta} \mid \mathcal{Y}) $
with a tractable surrogate distribution $q(\boldsymbol{\Theta})$ drawn from a restricted family of distributions $\EuScript{Q}$ that will be specified below. This approximation is obtained by minimizing the \ac{kl} divergence
\begin{equation}
\mathrm{KL}\!\big(q(\boldsymbol{\Theta}) \,\|\, p(\boldsymbol{\Theta} \mid \mathcal{Y}) \big)
=
\int q(\boldsymbol{\Theta})
\log \frac{q(\boldsymbol{\Theta})}{p(\boldsymbol{\Theta} \mid \mathcal{Y}) }
\, d\boldsymbol{\Theta}.
\end{equation}

Furthermore, minimizing the \ac{kl} divergence is equivalent to maximizing the \ac{elbo}~\cite{bishop2011Pattern}, defined as 
\begin{equation}
\label{eq:elbo}
\mathcal{L}(q) = \mathbb{E}_{q(\boldsymbol{\Theta})} \big[ \log p(\mathcal{Y}, \boldsymbol{\Theta}) \big] - \mathbb{E}_{q(\boldsymbol{\Theta})}
\big[ \log q(\boldsymbol{\Theta}) \big],
\end{equation}
and the goal is therefore to solve the maximization problem
\begin{equation}
 q^\star(\boldsymbol{\Theta}) = \argmax_{q(\boldsymbol{\Theta})\in\EuScript{Q}}\mathcal{L}(q).
\label{eq:opt_elbo}
\end{equation}

Under the mean-field approximation~\cite{Beal2003VariationalAF}, the variational distribution considers $\EuScript{Q}$ such that the latent variables distributions are independent, i.e., 
\begin{equation}
q(\boldsymbol{\Theta}) = q(\mathbf{C})\,q(\mathbf{K}) \, q(\mathbf{G}) \, q(\boldsymbol{\gamma}) \, q(\tau),
\end{equation}
thereby breaking the statistical dependencies present in the true posterior.
This factorization facilitates coordinate-wise unconstrained optimization over the posterior distribution, in which each variational variable is updated by taking the expectation of the joint log-density with respect to all other variables.
Specifically, the optimal solution of each variational component is
\begin{equation}
\log q^\star(\theta_i)
=
\mathbb{E}_{q(\boldsymbol{\Theta} \setminus \theta_i)}
\big[ \log p(\mathcal{Y}, \boldsymbol{\Theta}) \big]
+ \underline{\mathrm{c}},
\label{eq:mean_field}
\end{equation}
where $\theta_i$ denotes the $i$-th component of $\boldsymbol{\Theta} = \{\mathbf{C},\,\mathbf{K},\mathbf{G},\, \boldsymbol{\gamma},\, \tau\}$ for $i\in\{1,3,4,5\}$.
This yields a set of coupled yet tractable update equations that are iteratively solved until convergence. 
However, for $\theta_2=\mathbf{K}$, we consider further constraints on its approximate posterior distribution (hence on $\EuScript{Q}$) in order to maintain the underlying Khatri-Rao structure.
Specifically, since the constraint-free distribution using \eqref{eq:mean_field} for $\theta_2$ is a Gaussian distribution, we choose to consider the approximate posterior density function as
\begin{equation}
    q(\mathbf{K}) = \prod_{k=1}^{K_\mathrm{a}} \mathcal{CN}(\mathbf{k}_k ; \widehat{\mathbf{k}}_k, \delta \mathbf{I}_{T_K}),
\label{eq:cons_post}
\end{equation}
with the additional constraints that the variational mean $\widehat{\mathbf{k}}_k$ is characterized by $\widehat{\mathbf{k}}_k = \mathbf{h}_k \otimes \mathbf{a}_{k,d-1} 
\otimes \cdots \otimes \mathbf{a}_{k,2} \otimes \mathbf{a}_{k,1} $ and $\delta$ is a scalar variational parameter representing the shared posterior variance across all columns of $\mathbf{K}$. 
We additionally impose a second moment constraint on $\mathbf{K}$, motivated by its Khatri-Rao structure~\eqref{eq:kr}: the Grassmannian factors $\mathbf{A}_i$ have fixed column norms $\sqrt{T_i}$, and the channel factor $\mathbf{H}$ contributes expected energy $\mathbb{E}[\|\mathbf{h}_k\|^2] = N$ per column, so each column of $\mathbf{K}$ has expected squared norm $T_K = N\prod_i T_i$. We therefore constrain the total expected energy of $\mathbf{K}$ to $K_\text{a} T_K$. A similar constraint on $q(\mathbf{C})$ is not imposed explicitly, since the discrete mixture prior already pulls $q(\mathbf{C})$ toward the fixed-norm constellation points $\{\mathbf{v}_m\}$, and this pull becomes effective once the scale ambiguity is resolved by the constraint on $q(\mathbf{K})$. Since
\begin{equation}
    \mathbb{E}\!\big[\|\mathbf{k}_k\|^2\big] = \|\widehat{\mathbf{k}}_k\|^2 + \delta T_K,
\end{equation}
summing over all $K_\mathrm{a}$ columns, the total second moment is:
\begin{equation}
    \mathbb{E}\!\big[\normF{\mathbf{K}}^2\big] = \normF{\widehat{\mathbf{K}}}^2 + \delta K_\mathrm{a} T_K.
\end{equation}

We therefore impose the constraint:
\begin{equation}\label{eq:K_constraint}
    \mathbb{E}\!\big[\normF{\mathbf{K}}^2\big] \leq K_\mathrm{a} T_K
    \quad \Longleftrightarrow \quad
    \normF{\widehat{\mathbf{K}}}^2 + \delta K_\mathrm{a}T_K \leq K_\mathrm{a} T_K.
\end{equation}
Solving \eqref{eq:opt_elbo} over $q(\mathbf{K})$ then reduces to a constrained parametric search over $(\widehat{\mathbf{K}}, \widehat\delta)$, yielding tractable coordinate-ascent updates.

\subsection{Unconstrained Posterior Distribution Derivation}
Under the mean-field factorization, each variational factor is the exponential of the joint log-density averaged over the remaining factors, as in~\eqref{eq:mean_field}, reducing to isolating the terms that depend on the parameter of interest and propagating expectations through the moment identities of Appendix~\ref{app:vi}. The detailed derivations for each posterior are reported in Appendix~\ref{app:posteriors}.
Applying the mean-field approximation~\eqref{eq:mean_field}, expanding the joint distribution~\eqref{eq:joint}, and retaining only the terms that depend on each variational component, we can first deduce the family distributions of $\mathbf{C},\mathbf{G}, \boldsymbol{\gamma}, \tau$ and provide notations for their parameterizations.
\begin{itemize}[leftmargin=*, itemsep=4pt, topsep=4pt]
\item The approximate posterior on $\mathbf{C}$ is a matrix Gaussian distribution of mean $\widehat{\mathbf{C}}$ and covariance $\boldsymbol{\Sigma}_{\mathbf{C}}$.
\item The approximate posterior on $\mathbf{K}$ is chosen to be a matrix Gaussian distribution already defined in \eqref{eq:cons_post}.
\item The approximate posterior on $\mathbf{G}$ is a categorical distribution characterized by 
\begin{equation}
     \widehat{g}_k[m] \triangleq \mathbb{E}_{q(\mathbf{G})}[g_k[m]].
\end{equation}
\item The approximate posterior on $\boldsymbol{\gamma}$ is a product of independent Gamma distributions of parameters $\{\widehat{a}_{\gamma,k},\widehat{b}_{\gamma,k}\}$. We introduce the additional parameters
\begin{equation}
\widehat{\gamma}_k=\mathbb{E}_{q(\gamma_k)}[\gamma_k] = \frac{\widehat{a}_{\gamma,k}}{\widehat{b}_{\gamma,k}}.
\end{equation}
\item The approximate posterior on $\tau$ is a Gamma distribution with parameters $\widehat{a}_{\tau}$ and $\widehat{b}_{\tau}$. We introduce the additional parameter
\begin{equation}
\widehat{\tau}=\mathbb{E}_{q(\tau)}[\tau] = \frac{\widehat{a}_{\tau}}{\widehat{b}_{\tau}}.
\end{equation}
\end{itemize}

Let us now derive the solutions of \eqref{eq:opt_elbo} expressed with these parameters.

\emph{Posterior of $\mathbf{C}$.}
The derivations in Appendix~\ref{app:posteriors} provide
\begin{equation}
\log q(\mathbf{C})
=-\Tr\!\big(\mathbf{C}\,\mathbf{M}\,\mathbf{C}\hermconj\big)
+2\Re\!\big\{\Tr\!\big(\mathbf{C}\hermconj\mathbf{B}\big)\big\}+\underline{\mathrm{c}},
\label{eq:quad_form}
\end{equation}
where
\begin{equation}
\mathbf{M}=\widehat{\tau}\big(\widehat{\mathbf{K}}\trans\widehat{\mathbf{K}}^{*}+\widehat{\delta}T_K\mathbf{I}_{K_\mathrm{a}}\big)+\boldsymbol{\Sigma},
\label{eq:M_def}
\end{equation}
and
\begin{equation}
\mathbf{B}=\widehat{\tau}\mathbf{Y}^{(d)}\widehat{\mathbf{K}}^{*}+\widehat{\mathbf{V}}\boldsymbol{\Sigma},
\label{eq:B_def}
\end{equation}
with $\boldsymbol{\Sigma}=\text{diag}(\widehat\gamma_k)$.
Identifying~\eqref{eq:quad_form} with a Gaussian exponent (Appendix~\ref{app:posteriors}) shows that $q(\mathbf{C})$ has covariance 
\begin{equation}
\boldsymbol{\Sigma}_\mathbf{C}
=\mathbf{M}^{-1}
=\big(\widehat{\tau}(\widehat{\mathbf{K}}\trans\widehat{\mathbf{K}}^{*}+\widehat{\delta}T_K\mathbf{I}_{K_\mathrm{a}})+\boldsymbol{\Sigma}\big)^{-1},
\label{eq:qc_cov}
\end{equation}
and mean
\begin{equation}
\widehat{\mathbf{C}}
=\mathbf{B}\mathbf{M}^{-1}
=\big(\widehat{\tau}\mathbf{Y}^{(d)}\widehat{\mathbf{K}}^{*}+\widehat{\mathbf{V}}\boldsymbol{\Sigma}\big)\boldsymbol{\Sigma}_\mathbf{C}.
\label{eq:qc_mean}
\end{equation}
This estimate admits a direct least-squares interpretation: $\mathbf{M}$ is a regularized Gram matrix and $\mathbf{B}$ combines the likelihood term $\widehat{\tau}\mathbf{Y}^{(d)}\widehat{\mathbf{K}}^{*}$ with a penalty $\widehat{\mathbf{V}}\boldsymbol{\Sigma}$ that pulls each column of $\widehat{\mathbf{C}}$ toward the point $\widehat{\mathbf{v}}_k=\sum_m\widehat{g}_k[m]\mathbf{v}_m$, meaning that $\widehat{\mathbf{C}}=\mathbf{B}\mathbf{M}^{-1}$ is a regularized least-squares solution. The weight $\widehat{\tau}$ is the estimated noise precision, and $\boldsymbol{\Sigma}$ a diagonal matrix that collects the per-user weights $\widehat{\gamma}_k$. The term $\widehat{\delta}T_K\mathbf{I}_{K_\mathrm{a}}$ in $\mathbf{M}$ is the only one that contains the residual uncertainty $\widehat{\delta}$ of the estimated $\widehat{\mathbf{K}}$: when $\widehat{\mathbf{K}}$ is poorly resolved, it increases the regularization and biases $\widehat{\mathbf{C}}$ toward the constellation rather than overfitting the data.

\emph{Posterior of $\mathbf{G}$.}
Applying the mean-field approximation~\eqref{eq:mean_field}, expanding the joint distribution~\eqref{eq:joint}, and retaining only the terms that depend on the responsibilities $\mathbf{G}$ yields
\begin{equation}
\begin{aligned}
\log q(\mathbf{G})
&=\mathbb{E}_{q(\mathbf{C}),q(\boldsymbol{\gamma})}
\!\big[\log p(\mathbf{C}\mid\mathbf{G},\boldsymbol{\gamma})\big]\\
&\quad+\log p(\mathbf{G})+\underline{\mathrm{c}},
\end{aligned}
\label{eq:lnqg_split}
\end{equation}
where the first term is the expected mixture prior and the second is the categorical prior. These terms are linear in the indicators  $g_k[m]$, since $p(\mathbf{C}\mid\mathbf{G},\boldsymbol{\gamma})$ and $p(\mathbf{G})$ are products over the one-hot entries. Therefore, $q(\mathbf{G})$ is categorical and factorizes across users. Here, $\widehat{g}_k[m] \triangleq \mathbb{E}_{q(\mathbf{G})}[g_k[m]]$ is the parameter 
of the categorical posterior, while the indicators $g_k[m]\in\{0,1\}$ 
remain binary. Collecting the coefficient of each $g_k[m]$ and normalizing over the $M$ components yields the softmax responsibilities \cite{bishop2011Pattern}
\begin{equation}
\widehat{g}_k[m]
=\frac{\exp\!\big(-\widehat{\gamma}_k\|\widehat{\mathbf{c}}_k-\mathbf{v}_m\|^2+\log\rho_m\big)}
{\sum_{m'=1}^{M}\exp\!\big(-\widehat{\gamma}_k\|\widehat{\mathbf{c}}_k-\mathbf{v}_{m'}\|^2+\log\rho_{m'}\big)}.
\label{eq:gkm_up}
\end{equation}
As a result, when $\widehat{\gamma}_k$ is large, the softmax concentrates the distribution on the closest point, recovering the hard projection of~\eqref{eq:hard_form}; instead, when $\widehat{\gamma}_k$ is small, the distribution spreads across several points, deferring the decision. 

\emph{Posterior of $\boldsymbol{\gamma}$.}
Applying the mean-field approximation~\eqref{eq:mean_field}, expanding the joint distribution~\eqref{eq:joint}, and retaining only the terms that depend on the precisions $\boldsymbol{\gamma}$ yield
\begin{equation}
\log q(\boldsymbol{\gamma})=\mathbb{E}_{q(\mathbf{C}),q(\mathbf{G})}\left[\log p(\mathbf{C}\mid\mathbf{G},\boldsymbol{\gamma})\right] +\log p(\boldsymbol{\gamma})+\underline{\mathrm{c}},
\label{eq:lnqgamma_split}
\end{equation}
where the first term is the expected mixture prior and the second is the Gamma hyperprior. Each $\gamma_k$ enters only through a $\log\gamma_k$ term and a term that is linear in $\gamma_k$, which is exactly the form of a Gamma log-density. Hence $q(\boldsymbol{\gamma})$ is a product of independent Gamma factors, one per user, with shape $\widehat{a}_{\gamma,k}=a_{\gamma,k}+T_d$ and rate $\widehat{b}_{\gamma,k}=b_{\gamma,k}+T_d(\boldsymbol{\Sigma}_\mathbf{C})_{kk} +\sum_{m}\widehat{g}_k[m]\,\|\widehat{\mathbf{c}}_k-\mathbf{v}_m\|^2$. The required posterior mean is therefore
\begin{equation}
\widehat{\gamma}_k
=\frac{a_{\gamma,k}+T_d}
{b_{\gamma,k}+T_d(\boldsymbol{\Sigma}_\mathbf{C})_{kk}
+\sum_{m}\widehat{g}_k[m]\,\|\widehat{\mathbf{c}}_k-\mathbf{v}_m\|^2}.
\label{eq:gamm_up}
\end{equation}
The estimate $\widehat{\gamma}_k$ is an inverse variance: its denominator measures how far $\widehat{\mathbf{c}}_k$ lies from the constellation, plus the posterior variance $T_d(\boldsymbol{\Sigma}_\mathbf{C})_{kk}$ of the estimate itself. A $\widehat{\mathbf{c}}_k$ close to one constellation point gives a small denominator, hence a large $\widehat{\gamma}_k$ that tightens the prior around that point; an ambiguous or poorly estimated $\widehat{\mathbf{c}}_k$ keeps $\widehat{\gamma}_k$ small and the assignment uncertain.

\emph{Posterior of $\tau$.}
Applying the mean-field approximation~\eqref{eq:mean_field}, expanding the joint distribution~\eqref{eq:joint}, and retaining only the terms that depend on the noise precision $\tau$ yields 
\begin{equation}\label{eq:ln_tau}
\log q(\tau) = \mathbb{E}_{q(\mathbf{C}),q(\mathbf{K})} \left[\log p(\mathcal{Y}\mid\mathbf{C},\mathbf{K},\tau)\right]+\log p(\tau)+\underline{\mathrm{c}},
\end{equation}
where the first term is the expected likelihood and the second is the Gamma prior. As with $\boldsymbol{\gamma}$, $\tau$ enters only through a $\log\tau$ term and a linear term in $\tau$, which is the form of a Gamma log-density. Hence $q(\tau)$ is Gamma, with shape and rate
\begin{equation}
\widehat{a}_\tau=a_0+TN,
\qquad
\widehat{b}_\tau=b_0+\Xi_{\mathcal{Y}},
\label{eq:tau_ab}
\end{equation}
where $\Xi_{\mathcal{Y}}=\mathbb{E}_{q(\mathbf{C}),q(\mathbf{K})}[\normF{\mathbf{Y}^{(d)}-\mathbf{C}\mathbf{K}\trans}^2]$ is the expected reconstruction error (see \eqref{eq:xi_Y}). 
The posterior mean is therefore
\begin{equation}
\widehat{\tau}=\frac{\widehat{a}_\tau}{\widehat{b}_\tau}=\frac{a_0+TN}{b_0+\Xi_{\mathcal{Y}}}.
\label{eq:tau_up}
\end{equation}
Crucially, $\Xi_{\mathcal{Y}}$ retains the second-order terms $\widehat{\delta}T_K\mathbf{I}_{K_\mathrm{a}}$ and $T_d\boldsymbol{\Sigma}_\mathbf{C}$ from the posteriors of $\mathbf{K}$ and $\mathbf{C}$, so the residual is not underestimated by ignoring the uncertainty in the factors.

\subsection{Constrained Optimization on the Posterior of $\mathbf{K}$}

Considering the constrained Gaussian form of $q(\mathbf{K})$ defined in (\ref{eq:cons_post}), the optimization \eqref{eq:opt_elbo} becomes a maximization over the means of the posterior distribution $\widehat{\mathbf{k}}_k$ and the variance $\widehat\delta$. The optimal variational parameters for the factor $\mathbf{K}$, denoted as $\{\widehat{\mathbf{K}}, \widehat{\delta}\}$, are obtained by maximizing the isolated \ac{elbo} $\mathcal{L}(\widehat{\mathbf{K}}, \widehat\delta)$ defined in \eqref{fig:crit_1} under the prescribed structural constraints (see Appendix~\ref{app:elboK}):
\begin{subequations}\label{eq:optproblem}
    \begin{IEEEeqnarray}{rll}
        \{\widehat{\mathbf{K}},\widehat{\delta}\}
    &= \argmax_{\mathbf{K}, \delta}
    \mathcal{L}(\mathbf{K}, \delta) \\
    \text{s.t.}\quad
    & \widehat{\mathbf{K}} =  \widehat{\mathbf{H}} \odot \widehat{\mathbf{A}}_{d-1} \odot \widehat{\mathbf{A}}_{d-2} \odot \dots \odot \widehat{\mathbf{A}}_1,\\
    &\normF{\widehat{\mathbf{K}}}^2 + \delta \,K_\mathrm{a} \, T_K \leq K_\mathrm{a} \, T_K,\label{eq:constcouple} \\
    &\delta \geq 0.
    \end{IEEEeqnarray}
    \label{eq:max1}
\end{subequations}

\begin{figure*}[!t]
\begin{align}
\mathcal{L}(\widehat{\mathbf{K}}, \delta) = {}& \underline{\mathrm{c}}
- \widehat{\tau}\normF{ \mathbf{Y}^{(d)} - \widehat{\mathbf{C}}\widehat{\mathbf{K}}^{\top} }^2
- \widehat{\tau} T_d \Tr\!\big( \widehat{\mathbf{K}}\hermconj \widehat{\mathbf{K}}\,\boldsymbol{\Sigma}_\mathbf{C} \big)
- \frac{\normF{\widehat{\mathbf{K}}}^2}{\kappa} - \delta T_K \big( \widehat{\tau}(\normF{ \widehat{\mathbf{C}}}^2 + T_d \Tr(\boldsymbol{\Sigma}_\mathbf{C})) + \tfrac{K_{\rm a}}{\kappa} \big)
+ K_{\rm a} T_K \log \delta
\label{fig:crit_1}
\end{align}
\noindent\rule{\textwidth}{0.5pt}
\end{figure*}
 
Solving \eqref{eq:optproblem} in closed form is not possible due in particular to the non-convexity of the tensor decomposition. Similarly to \ac{als}, we alternatively solve the optimization in $\delta$ as well as each mode contained in $\mathbf{K}$. Moreover, since the norm constraint cannot be taken into account in closed-form as well, we first handle the unconstrained version of \eqref{eq:optproblem} in an \ac{als} manner, noting $\{\delta_u,\mathbf{K}_u\}$ the unconstrained solutions before eventually projecting the solution in order to respect the norm constraint.

First, the optimal variance $\delta_{\mathrm{u}}$ to the unconstrained problem is found by setting $\partial \mathcal{L}/\partial \delta = 0$:
\begin{equation}
\delta_{\mathrm{u}} = \frac{K_\mathrm{a}}{ \widehat{\tau}  (\normF{\widehat{\mathbf{C}}}^2 + T_d \text{Tr}(\boldsymbol{\Sigma}_\mathbf{C})) + \frac{K_\mathrm{a}}{\kappa}}.
\end{equation}

Second, rewriting the unconstrained optimization~\eqref{fig:crit_1} over $\mathbf{A}_i$ for $i=1,\dots, d+1$ with the additional notation $\mathbf{A}_{d+1}=\mathbf{H}$, leads to solving \eqref{eq:ai_crit}.
\begin{figure*}[!t]
\begin{equation}
\widehat{\mathbf{A}}_i
=
\arg\min_{\mathbf{A}_i}
\Bigg[
\widehat{\tau}
\normF{\mathbf{Y}^{(i)}-\mathbf{A}_i\mathbf{Z}_i\trans}^2
+
\mathrm{Tr}\Big(
(\mathbf{A}_i\hermconj\mathbf{A}_i * \mathbf{W}_i\hermconj\mathbf{W}_i)
\big(\widehat{\tau} T_d\boldsymbol{\Sigma}_C+\frac{\mathbf{I}_{K_\mathrm{a}}}{\kappa}\big)
\Big)
\Bigg].
\label{eq:ai_crit}
\end{equation}

\noindent\rule{\textwidth}{0.5pt}\nonumber
\end{figure*}

The solution $\mathbf{K}_{\mathrm{u}}
=
\widehat{\mathbf{H}}
\odot
\widehat{\mathbf{A}}_{d-1}
\odot
\cdots
\odot
\widehat{\mathbf{A}}_1$ is then defined through
\begin{equation}
\widehat{\mathbf{A}}_i =
\widehat{\tau} \mathbf{Y}_i \mathbf{Z}_i^*
\left[
\widehat{\tau} \mathbf{Z}_i^\top \mathbf{Z}_i^*
+
(\mathbf{W}_i\hermconj \mathbf{W}_i)*
\left(
\widehat{\tau} T_d \boldsymbol{\Sigma}_{\mathbf{C}} + \frac{\mathbf{I}_{K_\mathrm{a}}}{\kappa}
\right)
\right]^{-1},
\label{eq:ai_up}
\end{equation}
where
\begin{equation}
\mathbf{W}_i = \widehat{\mathbf{A}}_{d+1} \odot \widehat{\mathbf{A}}_{d-1} \odot \cdots \odot \widehat{\mathbf{A}}_{i+1} \odot \widehat{\mathbf{A}}_{i-1} \odot \cdots \odot \widehat{\mathbf{A}}_1,
\end{equation}
and 
\begin{equation}
\mathbf{Z}_i = \widehat{\mathbf{A}}_{d+1} \odot \widehat{\mathbf{C}} \odot \cdots \odot \widehat{\mathbf{A}}_{i+1} \odot \widehat{\mathbf{A}}_{i-1} \odot \cdots \odot \widehat{\mathbf{A}}_1.
\end{equation}

Finally, the norm constraint~\eqref{eq:constcouple} couples the mean and the variance of the approximate posterior distribution through the shared energy budget $K_\text{a}T_K$. If this budget is violated, we sequentially project each unconstrained estimate.
First, we consider the energy budget of $\delta$ with respect to the former estimate $\widehat{\mathbf{K}}^{(t)}$ (where $t$ refers to the iteration index of the variational inference) and reinject $\widehat{\mathbf{K}}^{(t)}$ in \eqref{eq:constcouple}. We obtain an upper bound on $\delta$ as 
\begin{equation}
    \delta_{\max} = \max\left(0,1 - (1+\eta)\frac{\normF{\widehat{\mathbf{K}}^{(t)}}^2}{K_\mathrm{a} T_K}\right),
\end{equation}
with $\eta>0$ a safety parameter. Hence,
\begin{equation}
\widehat{\delta}=
\begin{cases}
\delta_{\rm u}
& \text{if } \delta_{\rm u}<\delta_{\max}, \\
\delta_{\max},
& \text{otherwise}.
\end{cases}
\label{eq:delt}
\end{equation}
We then project $\widehat{\mathbf{K}}_{\rm u}$ onto the residual budget $C=K_\mathrm{a}T_K-\widehat{\delta}K_\mathrm{a}T_K$,
\begin{equation}
\widehat{\mathbf{K}} =
\begin{cases}
\widehat{\mathbf{K}}_{\rm u}\,\frac{\sqrt{C}}{\normF{\widehat{\mathbf{K}}_{\rm u}}},
& \text{if } \normF{\widehat{\mathbf{K}}_{\rm u}}^2 > C, \\
\widehat{\mathbf{K}}_{\rm u},
& \text{otherwise}.
\end{cases}
\label{eq:k_eq}
\end{equation}

We considered a slightly relaxed budget on $\delta$ (with $\eta$ typically $0.01$) rather than the strict full budget corresponding to $\eta=0$, since a strict projection can stall the $\delta$ update at a fixed point: if $\widehat{\delta}^{(t+1)}=1-\normF{\widehat{\mathbf{K}}^{(t)}}^2/(K_\mathrm{a}T_K)$, then the hard projection typically forces $\normF{\widehat{\mathbf{K}}^{(t)}}^2 = C = K_\mathrm{a}T_K-\widehat{\delta}^{(t)}K_\mathrm{a}T_K$,
which gives $\widehat{\delta}^{(t+1)}=\widehat{\delta}^{(t)}$. The variance then stagnates, preventing further updates; the small relaxation $\eta$ breaks this fixed point and restores convergence.

\begin{algorithm}[!t]
\caption{\ac{dvbals}}
\label{alg:bayesian_dtf}
\begin{algorithmic}[1]
\Require $\mathcal{Y}$, $\epsilon_{\mathrm{iter}}$, $J_{\max}$, $\kappa$
\Ensure $\widehat{\mathbf{A}}_1,\dots,\widehat{\mathbf{A}}_{d-1}, \widehat{\mathbf{H}}, \widehat{\mathbf{C}}$
\State Initialize $\widehat{\mathbf{A}}_i \forall i, \widehat{\mathbf{G}},\mathbf{\Sigma_C}, a_0, b_0, a_{\gamma,k}, b_{\gamma,k}$ for all $k$
\State Initialize $\widehat\tau=\frac{a_0}{b_0}$ and $\widehat\gamma_k=\frac{a_{\gamma,k}}{b_{\gamma,k}}$
\For{$t \gets 1$ to $J_{\max}$}
    \State Update $\widehat{\delta}$ via~\eqref{eq:delt}
    \State Update $\widehat{\mathbf{A}}_i\ \forall i$ and $\widehat{\mathbf{H}}$ via~\eqref{eq:ai_up}
    \State Compute $\widehat{\mathbf{K}}$ via~\eqref{eq:k_eq}
    \State Update $\widehat{\boldsymbol{\Sigma}}_{\mathbf{C}}$ and $\widehat{\mathbf{C}}$ via~\eqref{eq:qc_cov} and~\eqref{eq:qc_mean}, respectively.
    \State Update $\widehat{\gamma}_k\ \forall k$ via~\eqref{eq:gamm_up}
    \State Update $\widehat{\mathbf{g}}_k\ \forall k$ via~\eqref{eq:gkm_up}
    \State Update $\widehat{\tau}$ via~\eqref{eq:tau_up}
   \If {$\normF{\boldsymbol{\Psi}^{(t)} - 
\boldsymbol{\Psi}^{(t-1)}} < \epsilon_{\mathrm{iter}}
\normF{\boldsymbol{\Psi}^{(t-1)}}$}
\State break
\EndIf

\EndFor
\State \Return $\widehat{\mathbf{A}}_1,\dots,\widehat{\mathbf{A}}_{d-1}, \widehat{\mathbf{H}}, \widehat{\mathbf{C}}$
\end{algorithmic}
\end{algorithm}
Algorithm~\ref{alg:bayesian_dtf} summarizes the proposed \ac{dvbals} procedure. The algorithm takes as input tensor $\mathcal{Y}$, the convergence threshold $\epsilon_{\mathrm{iter}}$, the maximum number of iterations $J_{\max}$ and the hyperparameter $\kappa$, and returns the tensor factors $\widehat{\mathbf{A}}_1,\dots,\widehat{\mathbf{A}}_{d-1}, \widehat{\mathbf{H}}, \widehat{\mathbf{C}}$. Upon initialization, which is detailed in Section~\ref{sec:num_res}, the algorithm runs iteratively until either $J_{\max}$ or the convergence threshold is reached. At each iteration, the algorithm performs closed-form coordinate-ascent updates for all latent variables. Convergence is monitored via the relative change of the collection of estimated factor matrices $\boldsymbol{\Psi}^{(t)} \triangleq \{\widehat{\mathbf{A}}_i^{(t)}, \widehat{\mathbf{H}}^{(t)}, \widehat{\mathbf{C}}^{(t)}\}_{i=1}^{d-1}$, where $\|\boldsymbol{\Psi}^{(t)}\|_{\rm F}^2 \triangleq \sum_{i=1}^{d-1} \|\widehat{\mathbf{A}}_i^{(t)}\|_{\rm F}^2 + \|\widehat{\mathbf{H}}^{(t)}\|_{\rm F}^2 + \|\widehat{\mathbf{C}}^{(t)}\|_{\rm F}^2$. The algorithm terminates when $\|\boldsymbol{\Psi}^{(t)} -\boldsymbol{\Psi}^{(t-1)}\|_{\rm F} < \epsilon_{\mathrm{iter}}\,\|\boldsymbol{\Psi}^{(t-1)}\|_{\rm F}$ .

\section{Discrete Bayesian Variational TBM Receiver}
\label{sec:dvbtbm}

\begin{figure}[t]
    \centering
    \includegraphics[width=1\linewidth]{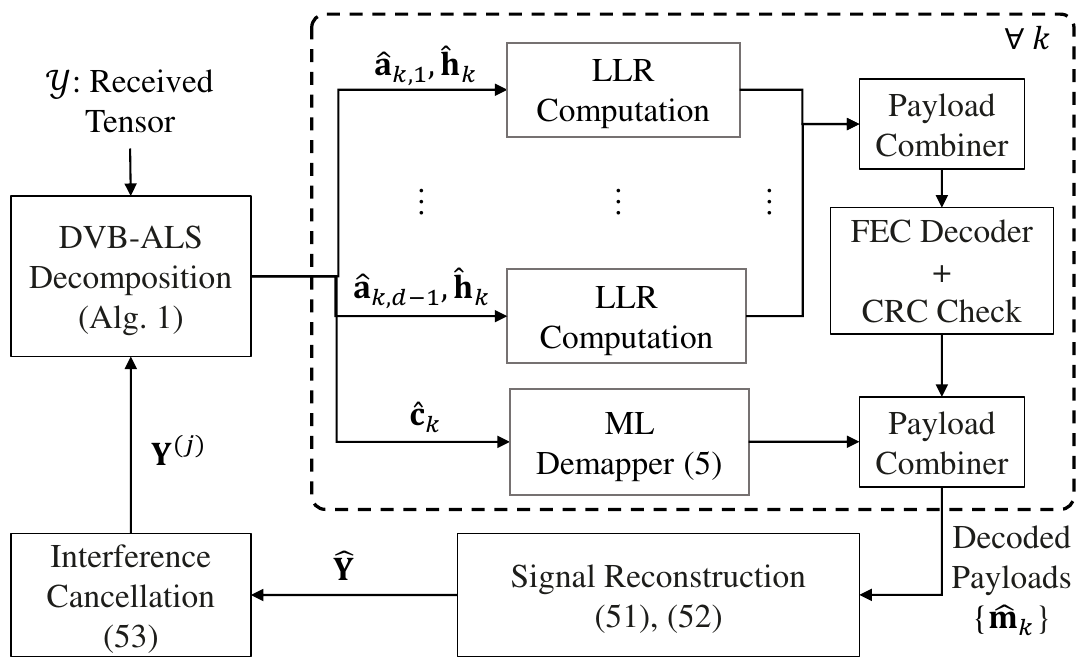}
        \caption{\ac{dvbtbm} decoder architecture.}
    \label{fig:dec}
\end{figure}

In this section, we present the proposed \ac{dvbtbm} receiver, which integrates \ac{dvbals} for user separation and combines it with a single-user demapper and \ac{sic}. The Bayesian \ac{cpd} stage provides estimates of the tensor factors associated with the active users, which are subsequently exploited to perform per-factor soft-demapping through \ac{llr} computation. The resulting soft information is passed to the channel decoder for message recovery, and successfully decoded users are iteratively removed from the received tensor through \ac{sic} to improve the detection of the remaining users. Since $\mathbf{c}_k$ does not originate from channel encoding, it is demapped from the codebook via a simple maximum a posteriori decision. The \ac{llr} computation is performed for each estimated factor $\widehat{\mathbf{a}}_{k,i}$ at the output of the user separation step similarly to \cite{Rech2023Unsourced}.
The \ac{llr} computed across all factors are then concatenated and fed to the \ac{fec} decoder with \ac{crc} verification. Successfully decoded sequences are then passed to a \ac{sic} step, where the corresponding contributions are removed from the received tensor, allowing the remaining users' signals to be decoded iteratively. The \ac{sic} step at the $j$-th iteration proceeds as follows: successfully decoded sequences are re-encoded to form the matrix
\begin{equation}
\overline{\mathbf{X}}^{(j)} \triangleq 
\left[\overline{\mathbf{x}}_1, \cdots, 
\overline{\mathbf{x}}_{|\EuScript{S}^{(j)}|}\right] 
\in \mathbb{C}^{T \times |\EuScript{S}^{(j)}|},
\end{equation}
where $\EuScript{S}^{(j)}$ denotes the set of successfully decoded sequences at iteration $j$ and $\overline{\mathbf{x}}_k \in \mathbb{C}^T$ is the re-encoded sequence of user $k$. The channel matrix associated with the decoded users is then estimated via least squares as
\begin{equation}
\widehat{\mathbf{H}}^{(j)} =
\left({\overline{\mathbf{X}}{^{(j)}}\hermconj}
\overline{\mathbf{X}}^{(j)}\right)^{-1}
\overline{\mathbf{X}}{^{(j)}}\hermconj\mathbf{Y}^{(j)},
\end{equation}
and the residual received signal is updated by subtracting the contribution of the decoded users as
\begin{equation}
\mathbf{Y}^{(j)} = \mathbf{Y}^{(j-1)} -\widehat{\mathbf{Y}},
\end{equation}
where  $\widehat{\mathbf{Y}}=
\sum_{k=1}^{|\EuScript{S}^{(j)}|}\overline{\mathbf{x}}_k \widehat{\mathbf{h}}_{k}\trans$ and $\widehat{\mathbf{h}}_{k}$ is the estimated channel vector of user $k$. The full \ac{dvbtbm} decoder architecture is summarized in Fig.~\ref{fig:dec}.

\section{Numerical Results}
\label{sec:num_res}

This section reports the numerical results, providing an extensive performance evaluation of \ac{dvbals} under different hyperparameter choices, as well as a comparison of \ac{dvbtbm} with the state-of-the-art \ac{ura} schemes.

\subsection{Performance Metrics}
\label{subsec:metrics}

We first evaluate the tensor decomposition performance as a function of the \ac{snr}, defined as $\mathrm{SNR}[\mathrm{dB}] = -10\log_{10}\sigma^2$, providing a clear assessment of factor-recovery accuracy independently of the single-user demapping step. To this end, we consider a miss-detection metric computed directly from the tensor factors estimated by the decomposition algorithms under comparison: conventional \ac{als}~\cite{Decurninge2021Tensor}, Hybrid-\ac{als}~\cite{Baccar2025Tensor}, and the proposed Bayesian decomposition. This metric is chosen to assess the quality of the recovered tensor factors on their own, before the influence of channel decoding.
For the $k$-th active user, the transmitted signal vector is given by equation \eqref{eq:xk}, while its estimate reconstructed from the \ac{cpd} output is $\widehat{\mathbf{x}}_k$. A miss-detection is declared for an estimated component $\widehat{\mathbf{x}}_k$ if its maximum normalized correlation with all transmitted vectors is below a predefined threshold $\beta$.
If multiple transmitted vectors yield correlations above the threshold, they are not distinguished, and no additional penalty is applied. The tensor decomposition miss-detection probability is then defined as
\begin{equation}
p_{\mathrm{md}}^{\mathrm{TD}} = \frac{1}{K_{\mathrm{a}}} \sum_{k=1}^{K_{\mathrm{a}}} \mathbb{I}
\left(
\max_{1 \le i \le K_{\mathrm{a}}}
\frac{\big| \widehat{\mathbf{x}}_k\hermconj \mathbf{x}_i \big|}
{\| \widehat{\mathbf{x}}_k \| \, \| \mathbf{x}_i \|}
< \beta
\right).
\end{equation}
Note that the metric $p_{\mathrm{md}}^{\mathrm{TD}}$ is not the probability of miss-detection of the \ac{ura} system; it is used only as a decomposition error metric for comparison purposes of the different tensor decomposition algorithms.

The performance of \ac{ura} schemes, instead, is measured by the receiver's decoding capability. The standard evaluation metric is the \emph{\ac{pupe}}, which measures the overall performance of the \ac{ura} scheme as
\begin{equation}
    p_{\mathrm{e}} = \min\left\{\mathbb{E}\left[\frac{|\EuScript{L}\setminus\widehat{\EuScript{L}}|}{K_{\rm a}}\right] + \mathbb{E}\left[\frac{|\widehat{\EuScript{L}}\setminus\EuScript{L}|}{|\widehat{\EuScript{L}}|}\right], \, 1 \right\},
\end{equation}
where the expectations denote the probabilities of miss-detection and false alarm, respectively. The performance is evaluated against the 
energy-per-bit to noise power spectral density ratio, given by 
$\frac{E_b}{N_0} = \frac{T}{B\,\sigma^2}$ where $B$ is the number of bits transmitted by each UE. We will compare the proposed approach \ac{dvbtbm} with the baseline schemes \ac{fasura}\cite{Gkagkos2022FASURA}, PTURA\cite{Fang2025Polar}, and Hybrid-TBM~\cite{Baccar2025Tensor}.

\subsection{Performance Evaluation of DVB-ALS}

The first setup considered is aligned with the one in~\cite{Meng2021Advanced}, consisting in 6 new radio numerology-0 resource blocks, which map to $T=1008$ time-frequency resources~\cite{3GPP2026Physical}. $N=8$ antennas are deployed at the receiver side. 
The tensor dimensions are $(T_1, T_2, T_3, T_4) = (14, 6, 3, 4)$, where the last mode is the discrete-optimized one, and is assigned $z = 3$ bits. The \ac{cpd} miss-detection threshold is set to $\beta = 0.8$.
The hyperparameters $a_0$, $b_0$, $a_{\gamma,k}$, and $b_{\gamma,k}$ are set to $10^{-6}$, yielding weakly informative  Gamma priors so that the posterior is driven mainly by the observed data~\cite{bishop2011Pattern}. 
In Algorithm~\ref{alg:bayesian_dtf}, $\boldsymbol{\Sigma}_{\mathbf{C}}$ is initialized to $\mathbf{I}$, the initial elements in $\widehat{\mathbf{A}}_i$ are drawn independently from $\mathcal{CN}(\mathbf{0}, \mathbf{I}_{K_\mathrm{a}})$ and $\widehat{\mathbf{G}}$ is initialized uniformly with $\widehat{g}_k[m] = 1/M$ for all $k, m$.

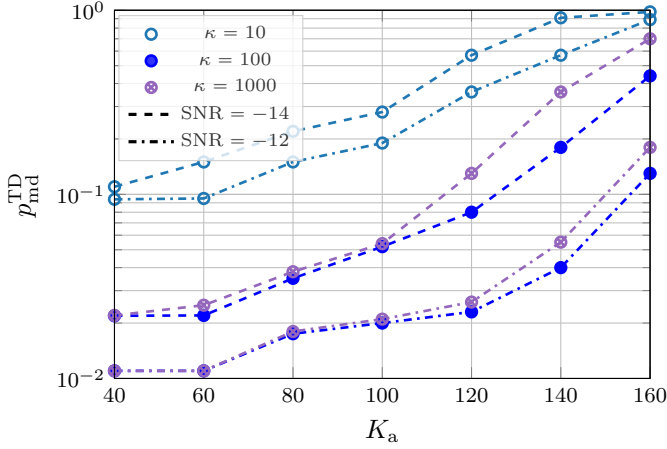
\begin{figure}
    \centering
    \setlength\fwidth{0.8\columnwidth}
    \setlength\fheight{0.55\columnwidth}
    \input{figs/pmdtd_vs_Ka_kappa_SNR}
    \caption{Tensor-decomposition miss-detection probability $p_{\mathrm{md}}^{\mathrm{TD}}$ versus $K_\mathrm{a}$ for different  \ac{snr} and $\kappa$ values.}
    \label{fig:pmdtd_vs_Ka_kappa_SNR}
\end{figure}
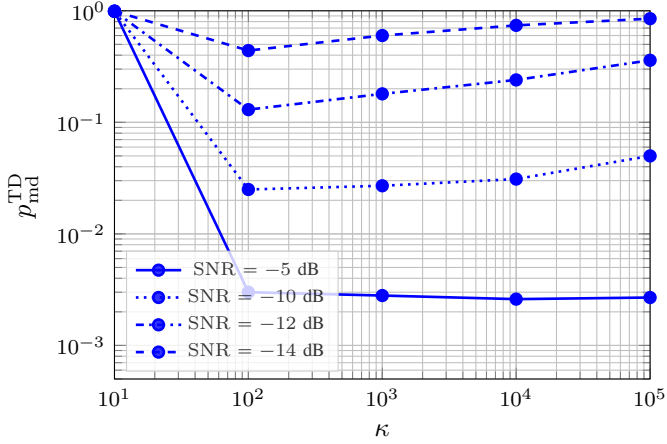
\begin{figure}
    \centering
    \setlength\fwidth{0.8\columnwidth}
    \setlength\fheight{0.55\columnwidth}
    \input{figs/pmdtd_vs_kappa}
    \caption{Tensor-decomposition miss-detection probability $p_{\mathrm{md}}^{\mathrm{TD}}$ versus the prior variance hyperparameter $\kappa$, for $K_\mathrm{a}=160$, $z=3$, and different \ac{snr} values.}
     \label{fig:pmdtd_vs_kappa}
\end{figure}

\emph{Sensitivity to the prior variance $\kappa$.} Since $\kappa$ controls the prior variance of $\mathbf{K}$ and is manually specified rather than inferred, we assess its effect through a sensitivity analysis over $300$ \ac{cpd} iterations. Fig.~\ref{fig:pmdtd_vs_Ka_kappa_SNR} reports the tensor decomposition misdetection probability $p_{\mathrm{md}}^{\mathrm{TD}}$ as a function of the number of active users $K_\mathrm{a}$ and for several values of $\kappa$. This figure shows that small values ($\kappa \leq 10$) cause complete detection failure regardless of \ac{snr} or user load, as over-regularization prevents the factor updates from recovering the signal. For larger $\kappa$, performance depends on the interplay between \ac{snr} and load: at low \ac{snr} ($-14$~dB), $\kappa = 100$ attains the lowest miss-detection probability across all loads, and its advantage is most pronounced at high loads where inter-user interference makes regularization more critical. When the \ac{snr} increases to $-12$~dB, the curves exhibit similar performance, while the performance gap between $\kappa=100$ and $\kappa=200$ is reduced.
Fig.~\ref{fig:pmdtd_vs_kappa} plots $p_{\mathrm{md}}^{\mathrm{TD}}$ directly against $\kappa$ for $K_\mathrm{a}=160$ and confirms that sensitivity to
$\kappa$ tends to vanish at higher \ac{snr}: at $\mathrm{SNR} = -5$~dB the performance is essentially flat starting from $\kappa\geq 100$. We therefore fix $\kappa = 100$ for this configuration, as it is shown to be the most robust choice across loads and \ac{snr} values, particularly in the demanding low-\ac{snr}, high-load regime, which is also the most relevant for \ac{ura}.

\begin{figure}
    \centering
    \setlength\fwidth{0.8\columnwidth}
    \setlength\fheight{0.55\columnwidth}
    \input{figs/pmdtd_vs_SNR_z}
    \caption{Tensor-decomposition miss-detection probability $p_{\mathrm{md}}^{\mathrm{TD}}$ versus \ac{snr}, for $K_\mathrm{a} = 160$, $\kappa = 100$, and different numbers of constellation bits $z$.}
     \label{fig:pmdtd_vs_SNR_z}
\end{figure}
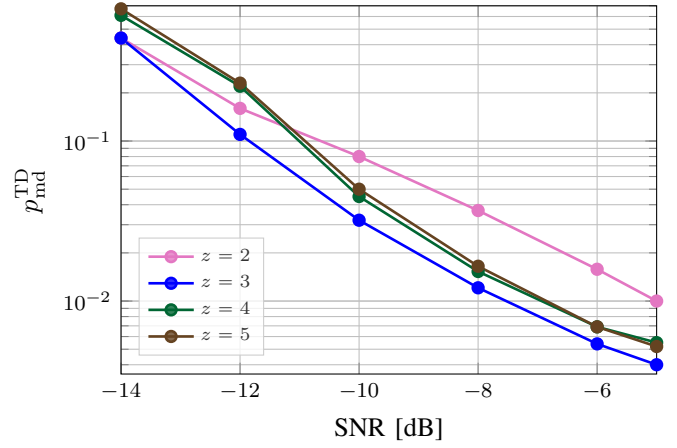

\emph{Sensitivity to the number of constellation bits $z$.} Recall that $z$ is the number of bits mapped to the $d$-th factor, setting the constellation size to $M = 2^z$. 
Fig.~\ref{fig:pmdtd_vs_SNR_z} shows that $z = 3$ achieves the best performance across all \ac{snr} values at $K_\mathrm{a}=160$, reflecting two competing effects: increasing $z$ enriches the sub-constellation diversity (hence $z = 2$ performs worst) at the cost of exponential complexity growth. However, large constellations make the variational posterior increasingly diffuse and less able to concentrate on the correct point. We therefore retain $z = 3$ throughout the rest of the paper.

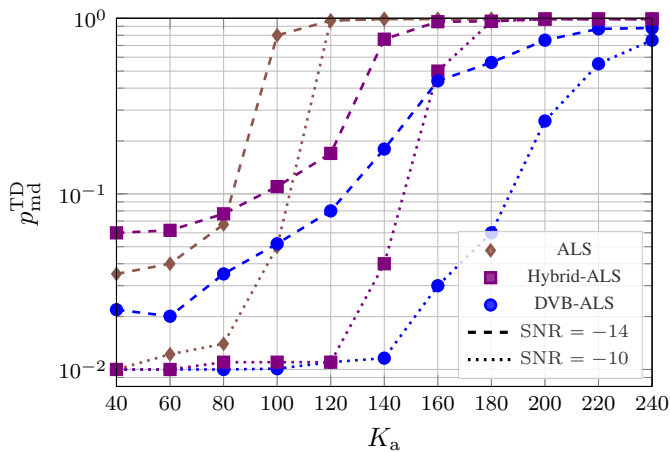
\begin{figure}
    \centering
    \setlength\fwidth{0.8\columnwidth}
    \setlength\fheight{0.55\columnwidth}
    \input{figs/pmdtd_vs_Ka_comp}
    \caption{Tensor-decomposition miss-detection probability $p_{\mathrm{md}}^{\mathrm{TD}}$ versus $K_\mathrm{a}$, for different \ac{snr} values.}
     \label{fig:pmdtd_vs_Ka_comp}
\end{figure}

\emph{Comparison of \ac{cpd} approaches.}
Fig.~\ref{fig:pmdtd_vs_Ka_comp} compares the proposed Bayesian \ac{cpd} against \ac{als} \cite{Kolda2009Tensor} and Hybrid-\ac{als}~\cite{Baccar2025Tensor} at $\mathrm{SNR} \in \{-14, -10\}$~dB, over a load range $40\leq K_\text{a}\leq 240$. Note that the theoretical maximum rank allowing complex tensor identifiability  in our setting is~\cite{Decurninge2021Tensor,Chiantini2014An} 
\begin{equation}
\overline{K}_\text{a} = \Bigg\lceil{\frac{TN}{N+\sum_{i=1}^d (T_i-1)}}\Bigg\rceil = 261.
\end{equation}
At both \ac{snr} values, \ac{als} degrades sharply beyond $K_\mathrm{a} = 100$. Hybrid-\ac{als} improves markedly by successfully separating more active users before saturation, but remains limited at low \ac{snr} ($-14$~dB), where the constellation projection becomes unreliable. The proposed scheme consistently outperforms both baselines across all loads and \ac{snr} values, maintaining stable performance well beyond the \ac{als} capacity limit by successfully exploiting the discrete nature of the last mode at the user separation step using the Bayesian framework.

\subsection{Performance Evaluation of DVB-TBM}
We now compare the proposed \ac{ura} scheme against state-of-the-art approaches, Hybrid-TBM, \ac{fasura}\cite{Gkagkos2022FASURA} and PTURA \cite{Fang2025Polar}. The evaluation metric is the energy efficiency, defined as the minimum $E_b/N_0$ ratio required to achieve a target per-user error probability $p_{\mathrm{e}} \leq 0.05$. 

We consider two configuration scenarios. For the first configuration set, we fix $T = 1008$ resources, a payload of $B = 110$ bits, and $N = 16$ antennas. A key constraint on the choice of the tensor encoding parameters considered in this paper is that only a few bits $z$ can be carried by the discrete mode, since increasing $z$ grows the constellation size exponentially as $2^z$, and hence the inference complexity, and the remaining payload must therefore be spread across the other modes. To balance this trade-off, we restrict the discrete mode to $T_d = 4$ ($z = 3$) and adopt the configuration $(T_1, T_2, T_3) = (18, 14, 4)$, where the first two modes carry $4$ bits per coordinate via the CubeSplit constellation \cite{Ngo2020Cube}, such that $\lfloor\log_2(T_i)\rfloor$ bits encode the face index and the remaining $(T_i-1)\times 4$ bits encode the local coordinates within that face. This yields $72$ bits for mode 1 and $55$ bits for mode 2, for a total encoded sequence of $127$ bits encoded with a polar code~\cite{Bioglio2021Design} and a \ac{crc}-aided list decoder of list size $12$. The discrete mode bits are not counted in the coded sequence since they bypass channel encoding and are decoded via direct maximum a posteriori decision. The payload $B = 110$ bits is chosen to expose a fundamental limitation of preamble-based \ac{ura} schemes like \ac{fasura}: as the load grows, its unsourced component~\cite{Gkagkos2023FASURA} incurs a high false-alarm rate that saturates the coherent part and prevents reliable user separation~\cite{Baccar2025Tensor}, an effect that worsens at large payloads. The proposed scheme instead uses the entire set of $T = 1008$ resources for both detection and decoding, inherently supporting higher loads and larger payloads, whereas \ac{fasura} must partition resources between its unsourced and coherent components, limiting its effective detection capacity. For a fair comparison, we use the best \ac{fasura} parameters for this configuration, i.e., $T^\mathrm{u} = 360$ resources and $B^\mathrm{u} = 14$ bits for the unsourced part, with the remaining resources allocated to the coherent part using a spreading sequence of length $4$. All the studied \ac{ura} schemes share the same number of \ac{sic} iterations which is set to $6$. Finally, we consider \ac{ptura}~\cite{Fang2025Polar} with the same tensor configuration as the one considered for the proposed scheme in which the three modes are encoded using the CubeSplit constellation. Thus, contrary to \ac{dvbtbm}, the last mode uses CubeSplit constellation, where $\lfloor\log_2(T_3)\rfloor$ bits encode the face index and the remaining $(T_3-1)\times 4$ bits encode the local coordinates within that face. The polar code rate is then adapted accordingly.
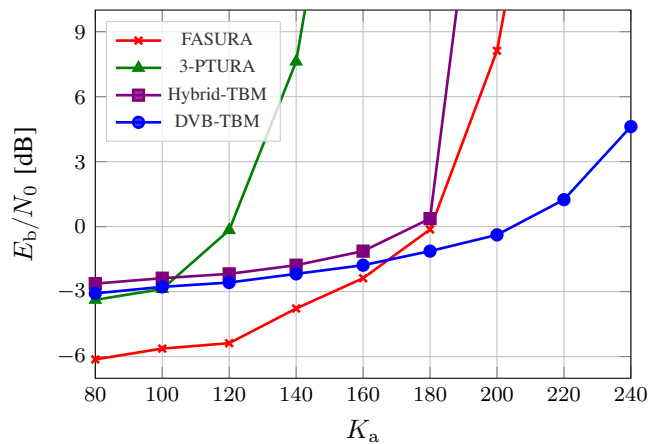
\begin{figure}
    \centering
    \setlength\fwidth{0.8\columnwidth}
    \setlength\fheight{0.55\columnwidth}
    \input{figs/ura_compa_N=16}
    \caption{Average minimum $E_b/N_0$ required to achieve $p_e \leq 0.05$ versus $K_{\rm a}$,
    for $N = 16$ and $T = 1008$.}
     \label{fig:ura1008res}
\end{figure}

Fig.~\ref{fig:ura1008res} reports the energy efficiency as a function of $K_\mathrm{a}$. At low \ac{snr}, associated with a small number of active users, \ac{fasura} performs better than the proposed scheme, which we attribute to the sub-optimality of the encoder, which maps only a small number of bits onto the discrete tensor mode. Moreover, 
\ac{fasura} operates effectively as a single-mode ($d=1$) scheme, granting 
it greater coding freedom than tensor-based approaches that split the 
payload across multiple modes and are therefore usually 
constrained to a higher code rate. This allows \ac{fasura} to employ more 
powerful coding and modulation schemes, with a lower code rate combined with random spreading\cite{Fang2025Polar}. However, at higher \ac{snr} and higher numbers of active users, \ac{fasura} saturates and fails to decode any user, while the proposed scheme remains robust and achieves reliable
detection across all tested loads. The proposed scheme also outperforms Hybrid-\ac{tbm}, which shares the same \ac{ura} decoder architecture but replaces the Bayesian \ac{cpd} block with Hybrid-\ac{als}; this confirms that the gain stems from the Bayesian tensor factorization itself rather than from the discrete decoder pipeline. 
Finally, \ac{ptura}~\cite{Fang2025Polar} degrades sharply at high loads, for two reasons. First, its Gaussian factor priors carry no information about the discrete sub-constellation structure, limiting identifiability. Second, its \ac{ard} mechanism severely underestimates the rank at high loads (a behavior visible in the authors' simulations results~\cite{Fang2025Polar}), likely because multi-user interference is absorbed into the noise estimate, causing the sparsity-promoting mechanism to prune active columns.

\begin{figure}
    \centering
    \setlength\fwidth{0.8\columnwidth}
    \setlength\fheight{0.55\columnwidth}
    \input{figs/ura_compa_N=50}
    \caption{Average minimum $E_b/N_0$ required to achieve $p_e \leq 0.05$ versus $K_{\rm a}$,
    for $N = 50$ and $T = 3200$.}
    \label{fig:ura3200res}
\end{figure}
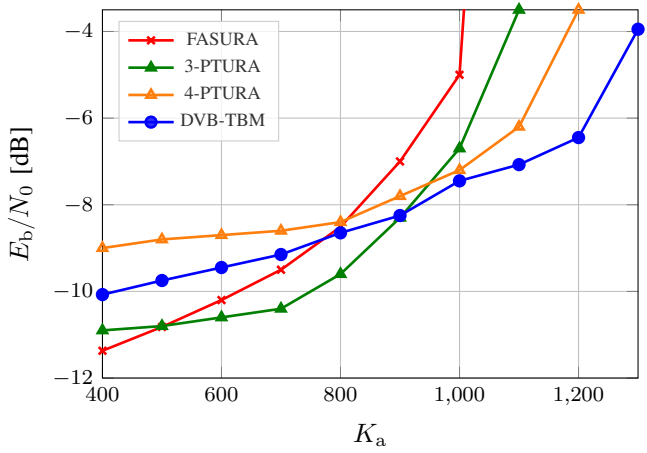

We finally evaluate the scheme under a second configuration, which is widely adopted in \ac{ura} literature\cite{Fengler2021SPARCs, Fengler21Pilot, Gkagkos2023FASURA, Decurninge2021Tensor, Fang2025Polar}: $T = 3200$ resources, $B = 96$ bits, and $N = 50$ antennas. We consider \ac{dvbtbm} with tensor configuration $(T_1, T_2, T_3) = (25, 16, 8)$, and $4$ bits encoded per each $\mathbf{a}_{k,i}$ coordinate in the first two modes and $z = 4$ in the discrete mode, yielding a \ac{fec} encoded sequence of $125$ bits. Furthermore, $\kappa$ was set to $300$ for this configuration. We compare our proposed scheme against \ac{fasura}~\cite{Gkagkos2023FASURA} and two variants of~\cite{Fang2025Polar}: 3-\ac{ptura} (three tensor modes) and 4-\ac{ptura} (four modes), shown in Fig.~\ref{fig:ura3200res}. At low \ac{snr}, the proposed scheme is weaker than \ac{fasura} and 3-\ac{ptura}, owing to the encoding strategy and in particular the choice to allocate $z = 4$ bits in the discrete mode. At higher \ac{snr} and dense loads, it surpasses both 3-\ac{ptura} and 4-\ac{ptura} in decoding capacity. This is notable because 4-\ac{ptura} uses an additional tensor mode yet is still outperformed by the proposed scheme with fewer modes, confirming the superior identifiability afforded by the discrete Bayesian prior.

\subsection{Computational Complexity Analysis}

\begin{table}
\centering
\caption{Computational Complexity of the Proposed Bayesian Algorithm Components}
\label{tab:complexity}
\begin{tabular}{l p{0.72\linewidth}}
\toprule
\textbf{Variable} & \textbf{Computational Complexity} \\
\midrule
$\widehat{\mathbf{A}}_i$, $\widehat{\mathbf{H}}$ & $\frac{1}{T_i} T K_\mathrm{a}^2 + TN K_\mathrm{a} + T_i K_\mathrm{a}^2 + K_\mathrm{a}^3$ \\
[4pt]
$\widehat{\mathbf{H}}$
& $T K_\mathrm{a}^2 +T K_\mathrm{a} + N K_\mathrm{a}^2 + K_\mathrm{a}^3$ \\[4pt]
$\widehat{\mathbf{C}}$
& $TN K_\mathrm{a} + T_d K_\mathrm{a}^2$ \\[4pt]
$\widehat{\boldsymbol{\Sigma}}_{\underline{\mathrm{c}}}$
& $\frac{T}{T_d} K_\mathrm{a}^2 + K_\mathrm{a}^3$ \\[4pt]
$\{\widehat g_k[m],\forall k\}$
& $K_\mathrm{a}M T_d$ \\[4pt]
$\{\widehat{\gamma}_k,\forall k\}$
& $K_\mathrm{a} M$ \\[4pt]
$\widehat{\tau}$
& $TN K_\mathrm{a} + T_d K_\mathrm{a}^2 + K_\mathrm{a}^3$ \\
\bottomrule
\end{tabular}
\end{table}

Table~\ref{tab:complexity} summarizes the per-iteration complexity of each variational update, expressed in terms of the number of complex multiplications (flops) with respect to a given tensor decomposition $(T_1,\dots,T_d)$. Notably, the factor updates $\widehat{\mathbf{A}}_i$ and $\widehat{\mathbf{H}}$ are dominated by the Gram matrix $\mathbf{Z}_i\hermconj\mathbf{Z}_i^*$ of cost $T K_\text{a}^2 / T_i$ flops and a $K_\text{a} \times K_\text{a}$ inversion of cost $K_\text{a}^3$ flops. Similarly, the update of $\widehat{\boldsymbol{\Sigma}}_C$ requires computing $\mathbf{K}\hermconj\mathbf{K}^*$ at cost $T K_\text{a}^2/T_d$ flops, followed by a $K_\text{a} \times K_\text{a}$ inversion. The discrete assignment $\{\widehat{g}_k \forall k\}$ requires $K_\text{a} M T_d$ flops, which remains tractable since $M = 2^z$ is kept small. Treating the tensor dimensions $T_1,\dots,T_d$ and the number of antennas $N$ as fixed and considering the scaling of the complexity with the number of active users $K_\text{a}$, the per-iteration multiplication complexity is dominated by the $\mathcal{O}(K_\text{a}^3)$ matrix inversions shared across multiple updates. Under this same regime, the proposed scheme shares the same per-iteration complexity scaling $\mathcal{O}(K_\text{a}^3)$ with the Bayesian CPD framework of~\cite{Fang2025Polar}, both being governed by the inversions required in the posterior mean updates. Furthermore, this complexity scaling is also shared with \ac{fasura}~\cite{Gkagkos2022FASURA}.

\section{Conclusion}
\label{sec:conclusions}
In this work, we proposed \ac{dvbals}, a variational Bayesian \ac{cpd} framework for \ac{tbm} in \ac{ura} that replaces the hard projection step of existing \ac{als}-based methods with a soft probabilistic alignment toward the sub-constellation points, obtained through a discrete Gaussian mixture prior on one factor matrix. The remaining factors are inferred through a structured Gaussian posterior whose mean is constrained to the Khatri-Rao product manifold. The result is a tractable closed-form coordinate ascent algorithm, which we integrated into \ac{dvbtbm}, a complete \ac{ura} receiver comprising single-user demapping, polar decoding with \ac{crc} verification, and \ac{sic}. Simulation results demonstrate significant gains over existing tensor decomposition methods and robust detection under the high system loads and payload sizes where state-of-the-art schemes saturate.

\appendices
\section{Derivation of the Variational Update Quantities}
\label{app:vi}
By the mean-field optimality condition~\eqref{eq:mean_field}, and substituting the joint factorization~\eqref{eq:joint}, every factor of $p(\mathcal{Y},\boldsymbol{\Theta})$ that does not contain $\theta_i$ contributes only an additive constant and is absorbed into the normalizer \cite{bishop2011Pattern}.

\emph{Log-densities.}
For a circularly symmetric complex Gaussian process of dimension $n$,
\begin{equation}
\begin{aligned}
\log\mathcal{CN}(\mathbf{x};\boldsymbol{\mu},\alpha^{-1}\mathbf{I}_n)
={}& n\log\alpha - n\log\pi \\
&- \alpha\,(\mathbf{x}-\boldsymbol{\mu})\hermconj(\mathbf{x}-\boldsymbol{\mu}),
\end{aligned}
\label{eq:lnCN}
\end{equation}
and for the Gamma density,
\begin{equation}
\log\mathcal{G}(x;a,b)=(a-1)\log x - b\,x + a\log b - \log\Gamma(a).
\label{eq:lnGamma}
\end{equation}

\emph{Moment of $\mathbf{K}$.}
The posterior $q(\mathbf{K})=\prod_{k}\mathcal{CN}( \mathbf{k}_k\mid\widehat{\mathbf{k}}_k,\widehat{\delta}\mathbf{I}_{T_K})$ has independent columns with $\mathbb{E}[\mathbf{k}_k]=\widehat{\mathbf{k}}_k$ and $\mathbb{E}[(\mathbf{k}_i-\widehat{\mathbf{k}}_i)(\mathbf{k}_j-\widehat{\mathbf{k}}_j)\hermconj]=\widehat{\delta}\,\delta_{ij}\mathbf{I}_{T_K}$.
Hence
\begin{equation}
\mathbb{E}_{q(\mathbf{K})}[\mathbf{K}\trans\mathbf{K}^{*}]
=\widehat{\mathbf{K}}\trans\widehat{\mathbf{K}}^{*}+\widehat{\delta}\,T_K\,\mathbf{I}_{K_\mathrm{a}}.
\label{eq:KK_moment}
\end{equation}

\emph{Moments of $\mathbf{C}$.}
The posterior $q(\mathbf{C})$ derived in Appendix~\ref{app:posteriors} has $T_d$ independent rows, each $\mathbf{c}_{r,:}\sim\mathcal{CN}(\widehat{\mathbf{c}}_{r,:},\boldsymbol{\Sigma}_\mathbf{C})$. Writing $\mathbf{c}_k$ for the $k$-th \emph{column} of $\mathbf{C}$, the per-element variance is $\mathbb{E}[|C_{rk}-\widehat{C}_{rk}|^2]=(\boldsymbol{\Sigma}_\mathbf{C})_{kk}$. Therefore,  summing over the $T_d$ rows yields

\begin{equation}
\begin{aligned}
\mathbb{E}_{q(\mathbf{C})}\!\big[\mathbf{C}\hermconj\mathbf{C}\big]
&=\widehat{\mathbf{C}}\hermconj\widehat{\mathbf{C}}
+\textstyle\sum_{r=1}^{T_d}\operatorname{Cov}(\mathbf{c}_{r,:})\\
&=\widehat{\mathbf{C}}\hermconj\widehat{\mathbf{C}}+T_d\boldsymbol{\Sigma}_\mathbf{C}.
\end{aligned}
\label{eq:Cmom_b}
\end{equation}

Furthermore, \begin{equation}
\begin{aligned}
\mathbb{E}_{q(\mathbf{C})}\!\big[\|\mathbf{c}_k-\mathbf{v}_m\|^2\big]
&=\|\widehat{\mathbf{c}}_k-\mathbf{v}_m\|^2
+\textstyle\sum_{r=1}^{T_d}\mathbb{E}[|C_{rk}-\widehat{C}_{rk}|^2]\\
&=\|\widehat{\mathbf{c}}_k-\mathbf{v}_m\|^2+T_d(\boldsymbol{\Sigma}_\mathbf{C})_{kk},
\end{aligned}
\label{eq:Cmom_a}
\end{equation}

\section{Derivation of the Unconstrained Variational Posterior Distribution}
\label{app:posteriors}

\emph{Posterior of $\mathbf{C}$.}
Applying the mean-field approximation~\eqref{eq:mean_field}, expanding the joint distribution~\eqref{eq:joint}, and retaining only the terms that depend on $\mathbf{C}$ yields
\begin{equation}\label{eq:mf_c}
\begin{aligned}
\log q(\mathbf{C})
&=
\mathbb{E}_{q(\mathbf{K}),q(\tau)}
\big[
\log p(\mathcal{Y} \mid \mathbf{C},\mathbf{K}, \tau)
\big] \\
&\quad+
\mathbb{E}_{q(\mathbf{G}) q(\boldsymbol{\gamma})}
\big[
\log p(\mathbf{C} \mid \mathbf{G}, \boldsymbol{\gamma})
\big]
+ \underline{\mathrm{c}},
\end{aligned}
\end{equation}
where the first term is the expected likelihood and the second is the mixture prior. Since the resulting log-density is a quadratic form in $\mathbf{C}$, it is the exponent of a Gaussian, so $q(\mathbf{C})$ is Gaussian.  Both components in \eqref{eq:mf_c} are quadratic in $\mathbf{C}$: the likelihood through its mean $\mathbf{C}\mathbf{K}\trans$, which is linear in $\mathbf{C}$, and the prior through the squared norm $\|\mathbf{c}_k-\mathbf{v}_m\|^2$. By~\eqref{eq:lnCN}, the log-likelihood is, up to constants, given by
\begin{equation}
\begin{aligned}
\log p(\mathcal{Y}\mid\mathbf{C},\mathbf{K},\tau)
={}&-\tau\normF{\mathbf{Y}^{(d)}-\mathbf{C}\mathbf{K}\trans}^2\\
={}&-\tau\Tr\!\big(\mathbf{C}\,\mathbf{K}\trans\mathbf{K}^{*}\,\mathbf{C}\hermconj\big)\\
&+2\tau\Re\{\Tr(\mathbf{C}\hermconj\mathbf{Y}^{(d)}\mathbf{K}^{*})\}+\underline{\mathrm{c}}.
\end{aligned}
\end{equation}
Substituting~\eqref{eq:KK_moment}, the expected likelihood term is
\begin{equation}
\begin{aligned}
&\mathbb{E}_{q(\mathbf{K}),q(\tau)}\!\big[\log p(\mathcal{Y}\mid\mathbf{C},\mathbf{K},\tau)\big] =\\
&\quad -\widehat{\tau}\Tr\!\Big(\mathbf{C}\big(\widehat{\mathbf{K}}\trans\widehat{\mathbf{K}}^{*}+\widehat{\delta}T_K\mathbf{I}_{K_\mathrm{a}}\big)\mathbf{C}\hermconj\Big)\\
&\quad +2\widehat{\tau}\Re\{\Tr(\mathbf{C}\hermconj\mathbf{Y}^{(d)}\widehat{\mathbf{K}}^{*})\}.
\end{aligned}
\label{eq:C_data}
\end{equation}
By~\eqref{eq:lnCN}, the log mixture prior keeping only $\mathbf{C}$-dependent terms is $-\sum_k\sum_m g_k[m]\gamma_k\|\mathbf{c}_k-\mathbf{v}_m\|^2$. Expanding the norm,
\begin{equation}
\|\mathbf{c}_k-\mathbf{v}_m\|^2
=\mathbf{c}_k\hermconj\mathbf{c}_k
-2\Re\{\mathbf{c}_k\hermconj\mathbf{v}_m\}
+\mathbf{v}_m\hermconj\mathbf{v}_m,
\end{equation}
the last term being $\mathbf{C}$-independent. The expectation $\mathbb{E}_{q(\mathbf{G}),q(\boldsymbol{\gamma})}[g_k[m]\gamma_k]$ factorizes as $\widehat{g}_k[m]\widehat{\gamma}_k$ because the mean-field family $q(\mathbf{G})q(\boldsymbol{\gamma})$ makes $g_k[m]$ and $\gamma_k$ independent, and $\mathbb{E}[g_k[m]]=\widehat{g}_k[m]$ for a one-hot indicator. Hence
\begin{equation}
\begin{aligned}
&\mathbb{E}\!\big[\log p(\mathbf{C}\mid\mathbf{G},\boldsymbol{\gamma})\big] = \underline{\mathrm{c}} -\\
&-\sum_k\widehat{\gamma}_k\Big[
\Big(\textstyle\sum_m\widehat{g}_k[m]\Big)\mathbf{c}_k\hermconj\mathbf{c}_k
-2\Re\Big\{\mathbf{c}_k\hermconj\textstyle\sum_m\widehat{g}_k[m]\mathbf{v}_m\Big\}\Big].
\end{aligned}
\end{equation}
where the one-hot constraint $\sum_m\widehat{g}_k[m]=1$ collapses the quadratic part and the definition $\widehat{\mathbf{v}}_k=\sum_m\widehat{g}_k[m]\mathbf{v}_m$ collapses the linear part. Stacking columns with $\boldsymbol{\Sigma}=\diag(\widehat{\gamma}_k)$ and $\widehat{\mathbf{V}}=[\widehat{\mathbf{v}}_1,\dots,\widehat{\mathbf{v}}_{K_\mathrm{a}}]$,
\begin{equation}
\begin{aligned}
\mathbb{E}\!\big[\log p(\mathbf{C}\mid\mathbf{G},\boldsymbol{\gamma})\big]
={}&-\Tr(\mathbf{C}\boldsymbol{\Sigma}\mathbf{C}\hermconj)\\
&+2\Re\{\Tr(\mathbf{C}\hermconj\widehat{\mathbf{V}}\boldsymbol{\Sigma})\}+\underline{\mathrm{c}}.
\end{aligned}
\label{eq:C_prior}
\end{equation}
Adding~\eqref{eq:C_data} and~\eqref{eq:C_prior} gives~\eqref{eq:quad_form}, with $\mathbf{M}$ and $\mathbf{B}$ as defined in~\eqref{eq:M_def} and ~\eqref{eq:B_def}, respectively. Since~\eqref{eq:quad_form} is quadratic in $\mathbf{C}$, it is a Gaussian log-density; matching it to the canonical Gaussian distribution
identifies the covariance as $\mathbf{M}^{-1}$ and the mean as the solution of $\mathbf{M}\boldsymbol{\mu}=\mathbf{B}$. 

\emph{Posterior of $\mathbf{G}$.}
From~\eqref{eq:joint}, the factors containing $\mathbf{G}$ are the mixture prior $p(\mathbf{C}\mid\mathbf{G},\boldsymbol{\gamma})$ and the categorical prior $p(\mathbf{G})=\prod_k\prod_m\rho_m^{g_k[m]}$, leading to~\eqref{eq:lnqg_split}.
By~\eqref{eq:lnCN} with $\alpha=\gamma_k$, $n=T_d$, the log mixture prior is
\begin{equation}
\begin{aligned}
\log p(\mathbf{C}\mid\mathbf{G},\boldsymbol{\gamma})
=\sum_k\sum_m g_k[m]\Big[
&T_d\log\gamma_k-T_d\log\pi\\
&-\gamma_k\|\mathbf{c}_k-\mathbf{v}_m\|^2\Big],
\end{aligned}
\end{equation}
and 
\begin{equation}
\log p(\mathbf{G})=\sum_k\sum_m g_k[m]\log\rho_m.
\end{equation} 
Both are linear in the indicators $g_k[m]$. Taking $\mathbb{E}_{q(\mathbf{C}),q(\boldsymbol{\gamma})}$ and using $\mathbb{E}[\log\gamma_k]=\psi(\widehat{a}_{\gamma,k})-\log\widehat{b}_{\gamma,k}$, $\mathbb{E}[\gamma_k]=\widehat{\gamma}_k$, together with the variance-corrected distance~\eqref{eq:Cmom_a}, gives
\begin{equation}
\begin{aligned}
\log q(\mathbf{G})
=\sum_k\sum_m g_k[m]\Big[
&T_d\big(\psi(\widehat{a}_{\gamma,k})-\log\widehat{b}_{\gamma,k}\big)\\
&-T_d\log\pi+\log\rho_m\\
&-\widehat{\gamma}_k\big(T_d(\boldsymbol{\Sigma}_\mathbf{C})_{kk}\\
&\qquad+\|\widehat{\mathbf{c}}_k-\mathbf{v}_m\|^2\big)
\Big]+\underline{\mathrm{c}}.
\end{aligned}
\label{eq:lnqg_full}
\end{equation}
Since $q(\mathbf{G})$ is categorical, $\widehat{g}_k[m]\propto\exp(\cdot)$ of the bracketed coefficient \cite{bishop2011Pattern}. The terms not depending on $m$ cancel with normalization. We collect the surviving $m$-dependent terms into the log-weight $-\widehat{\gamma}_k\|\widehat{\mathbf{c}}_k-\mathbf{v}_m\|^2+\log\rho_m$, so that normalizing over the $M$ components yields the softmax~\eqref{eq:gkm_up}.

\emph{Posterior of $\boldsymbol{\gamma}$.}
From~\eqref{eq:joint}, the factors containing $\boldsymbol{\gamma}$ are the mixture prior $p(\mathbf{C}\mid\mathbf{G},\boldsymbol{\gamma})$ and the Gamma hyperprior $p(\boldsymbol{\gamma})$, so ~\eqref{eq:lnqgamma_split}.
By~\eqref{eq:lnCN} with $\alpha=\gamma_k$, $n=T_d$, the log mixture prior is
\begin{equation}
\begin{aligned}
\log p(\mathbf{C}\mid\mathbf{G},\boldsymbol{\gamma})
=\sum_k\sum_m g_k[m]\Big[
&T_d\log\gamma_k-T_d\log\pi\\
&-\gamma_k\|\mathbf{c}_k-\mathbf{v}_m\|^2\Big].
\end{aligned}
\end{equation}
Taking $\mathbb{E}_{q(\mathbf{C}),q(\mathbf{G})}$, we use the variance-corrected distance~\eqref{eq:Cmom_a}. The $\log\gamma_k$ term simplifies through the one-hot constraint as $
\sum_m\widehat{g}_k[m]\,T_d\log\gamma_k
=T_d\log\gamma_k$. By~\eqref{eq:lnGamma}, the hyperprior contributes $(a_{\gamma,k}-1)\log\gamma_k-b_{\gamma,k}\gamma_k$. The distribution factorizes over $k$, and collecting the $\gamma_k$-dependent terms gives
\begin{equation}
\begin{aligned}
\log q(\gamma_k)
={}&\big(a_{\gamma,k}+T_d-1\big)\log\gamma_k\\
&-\gamma_k\Big[b_{\gamma,k}+T_d(\boldsymbol{\Sigma}_\mathbf{C})_{kk}\\
&\qquad+\sum_m\widehat{g}_k[m]\|\widehat{\mathbf{c}}_k-\mathbf{v}_m\|^2\Big]+\underline{\mathrm{c}},
\end{aligned}
\label{eq:lnqgamma_full}
\end{equation}
which is the log of a Gamma density. 

\emph{Posterior of $\tau$.}
From~\eqref{eq:joint}, the factors containing $\tau$ are the likelihood $p(\mathcal{Y}\mid\mathbf{C},\mathbf{K},\tau)$ and the Gamma prior $p(\tau)$ gives (\ref{eq:ln_tau}). By~\eqref{eq:lnCN} with $n=TN$,
\begin{equation}
\begin{aligned}
\log p(\mathcal{Y}\mid\mathbf{C},\mathbf{K},\tau)
={}&TN\log\tau-TN\log\pi\\
&-\tau\normF{\mathbf{Y}^{(d)}-\mathbf{C}\mathbf{K}\trans}^2.
\end{aligned}
\end{equation}
Therefore, collecting the $\tau$-dependent terms gives
\begin{equation}
\log q(\tau)=(a_0+TN-1)\log\tau-\tau\big(b_0+\Xi_{\mathcal{Y}}\big)+\underline{\mathrm{c}}.
\label{eq:lnqtau_full}
\end{equation}
Equation~\eqref{eq:lnqtau_full} is the log of a Gamma density; matching it to~\eqref{eq:lnGamma} gives $\widehat{a}_\tau=a_0+TN$ and $\widehat{b}_\tau=b_0+\Xi_{\mathcal{Y}}$, so $\widehat{\tau}=\widehat{a}_\tau/\widehat{b}_\tau$, which is~\eqref{eq:tau_up}.
To evaluate $\Xi_{\mathcal{Y}}$, we expand the squared norm as 
\begin{equation}
\begin{aligned}
\normF{\mathbf{Y}^{(d)}-\mathbf{C}\mathbf{K}\trans}^2
={}&\Tr(\mathbf{Y}^{(d)}\mathbf{Y}^{(d)\mathsf{H}})\\
&-2\Re\{\Tr(\mathbf{Y}^{(d)}\mathbf{K}^{*}\mathbf{C}\hermconj)\}\\
&+\Tr(\mathbf{K}\trans\mathbf{K}^{*}\mathbf{C}\hermconj\mathbf{C}),
\end{aligned}
\end{equation}
where the cyclic property is used for the quadratic term: $\Tr(\mathbf{C}\mathbf{K}\trans\mathbf{K}^{*}\mathbf{C}\hermconj) =\Tr(\mathbf{K}\trans\mathbf{K}^{*}\mathbf{C}\hermconj\mathbf{C})$. By taking the expectation, we obtain
\begin{equation}\label{eq:xi_Y}
\begin{aligned}
\Xi_{\mathcal{Y}}
={}&\Tr(\mathbf{Y}^{(d)}\mathbf{Y}^{(d)\mathsf{H}})
-2\Re\{\Tr(\mathbf{Y}^{(d)}\widehat{\mathbf{K}}^{*}\widehat{\mathbf{C}}\hermconj)\}\\
&+\Tr\!\Big((\widehat{\mathbf{K}}\trans\widehat{\mathbf{K}}^{*}+\widehat{\delta}T_K\mathbf{I}_{K_\mathrm{a}})(\widehat{\mathbf{C}}\hermconj\widehat{\mathbf{C}}+T_d\boldsymbol{\Sigma}_\mathbf{C})\Big).
\end{aligned}
\end{equation}

\section{Proof of \eqref{fig:crit_1}}
\label{app:elboK}
We begin from the variational objective in~\eqref{eq:joint} and consider only the terms that depend on $\mathbf{K}$ from the joint distribution~\eqref{eq:joint}.
Using~\eqref{eq:cons_post}, the entropy term simplifies to:
\begin{equation}
-\mathbb{E}_{q(\mathbf{K})}[\log q(\mathbf{K})]
= K_{\rm a} T_K (\log(\pi \widehat\delta) + 1).
\end{equation}
The expected quadratic reconstruction error is given by:
\begin{equation}
\begin{split}
\mathbb{E}_{q(\mathbf{C}),q(\mathbf{K}),q(\tau)}
\!\big[\normF{\mathbf{Y}^{(d)}-\mathbf{C}\mathbf{K}\trans}^2\big]
= -\widehat{\tau}\big(
\normF{\mathbf{Y}^{(d)}-\widehat{\mathbf{C}}\widehat{\mathbf{K}}\trans}^2 \\
+\, T_d\,\mathrm{Tr}(
\widehat{\mathbf{K}}\hermconj
\widehat{\mathbf{K}}
\boldsymbol{\Sigma}_{\mathbf{C}})
\\
+\, \widehat\delta T_K \big(
\normF{\widehat{\mathbf{C}}}^2
+ T_d\,\mathrm{Tr}(\boldsymbol{\Sigma}_{\mathbf{C}})
\big)
\big).
\end{split}
\end{equation}
By adding the prior term in~\eqref{eq:k_prior}, we obtain the combined objective:
\begin{equation}
\begin{aligned}
\mathcal{L}(\widehat{\mathbf{K}}, \widehat\delta) =
&-\widehat{\tau} \big(
\normF{ \mathbf{Y}^{(d)} - \widehat{\mathbf{C}}\widehat{\mathbf{K}}\trans}^2
+ T_d \text{Tr}(\widehat{\mathbf{K}}\hermconj \widehat{\mathbf{K}} \boldsymbol{\Sigma}_C)
\big) \\
&- \frac{1}{\kappa} \normF{ \widehat{\mathbf{K}}}^2- \widehat\delta T_K \bigg(
\widehat{\tau} (\normF{ \widehat{\mathbf{C}}}^2 + T_d \text{Tr}(\boldsymbol{\Sigma}_C))
+ \frac{K_\mathrm{a}}{\kappa}
\bigg) \\
&+ K_\mathrm{a}T_K \log \widehat\delta + \underline{\mathrm{c}}.
\end{aligned}
\end{equation}
Finally, substituting the Khatri-Rao structure defined in~\eqref{eq:kr} leads to the criterion in~\eqref{fig:crit_1}.

\bibliographystyle{IEEEtran}
\bibliography{IEEEabrv,biblio_abbrv}
\balance

\end{document}

%% file: figs/graph_model.tex
\begin{tikzpicture}[x=1.1cm, y=1cm, font=\footnotesize]

\node[const] (ag) {$a_{\gamma,k}$};
\node[const, right=0.4cm of ag] (bg) {$b_{\gamma,k}$};
\node[const, right=0.4cm of bg] (rho) {$\rho_m$}; 
\node[const, right=2.0cm of bg] (a0) {$a_0$}; 
\node[const, right=0.4cm of a0] (b0) {$b_0$};
\node[const, right=1.2cm of b0] (kappa) {$\kappa$}; 

\node[latent, below=0.8cm of ag, xshift=0.3cm] (gamma) {$\gamma_k$};
\node[latent, right=0.6cm of gamma] (G) {$\mathbf{g}_k$};

\node[latent, below=1.1cm of G, xshift=0.8cm] (ck) {$\mathbf{c}_k$};
\node[latent, below=0.8cm of a0, xshift=0.2cm] (tau) {$\tau$}; 
\node[latent, below=1.1cm of kappa] (K) {$\mathbf{K}$}; 

\node[obs, below=1.2cm of ck, xshift=1.6cm] (Y) {$\mathcal{Y}$};

\edge{ag}{gamma}
\edge{bg}{gamma}
\edge{a0}{tau}
\edge{b0}{tau}
\edge{kappa}{K}
\edge{rho}{G}

\edge{G}{ck}
\edge{gamma}{ck}
\edge{ck}{Y}
\edge{tau}{Y}
\edge{K}{Y}

\plate{plateK}{(gamma)(G)(ck)}{$k=1,\ldots,K_\mathrm{a}$};

\end{tikzpicture}

%% file: figs/pmdtd_vs_Ka_kappa_SNR.tex
\definecolor{colorDVBTBM}{RGB}{0, 0, 255}
\definecolor{colorkappa10}{RGB}{31, 119, 180}
\definecolor{colorkappa1000}{RGB}{148, 103, 189}  

\pgfplotsset{
    every tick label/.append style={font=\footnotesize},
    every axis plot/.append style={
        line width=1pt, 
        mark size=2,        
        mark options={solid}
    }
}

\tikzset{
    styleDVBTBM/.style={color=colorDVBTBM, mark=*},
    stylekappa10/.style={color=colorkappa10, mark=o},
    stylekappa1000/.style={color=colorkappa1000, mark=otimes},
}

\pgfplotsset{every tick label/.append style={font=\footnotesize}}
\begin{tikzpicture}
\begin{axis}[
    width=\fwidth,
    height=\fheight,
    at={(0\fwidth,0\fheight)},
    scale only axis,
    ylabel style={font=\normalsize},
    xlabel style={font=\normalsize},
    xlabel={$K_{\rm a}$},
    ylabel={$p_{\rm md}^{\rm TD}$},
    axis background/.style={fill=white},
    legend style={
      fill opacity=0.7,
      text opacity=1,
      at={(0.97,0.03)},
      anchor=south east,
      legend cell align=left, 
      align=left,
      font=\footnotesize,
      draw=lightgray204
    },
    xmajorgrids,
    xminorgrids,
    ymajorgrids,
    yminorgrids,
    xmin=40,
    xmax=160,
    ymin=0.01, 
    ymax=1,
    ymode=log,
]



\addplot [stylekappa10, dashed]
table {%
40 0.11
60 0.15
80 0.22
100 0.28
120 0.57
140 0.91
160 0.98
};
\addlegendentry{$\kappa = 10$ SNR =-14}

\addplot [styleDVBTBM, dashed]
table {%
40 0.0219
60 0.022
80 0.035
100 0.052
120 0.08
140 0.18
160 0.44
};
\addlegendentry{$\kappa  = 100$ SNR =-14}

\addplot [stylekappa1000, dashed]
table {%
40 0.022
60 0.025
80 0.038
100 0.054
120 0.13
140 0.36
160 0.7
};
\addlegendentry{$\kappa  = 1000$ SNR =-14}



\addplot [stylekappa10, dashdotted]
table {%
40 0.094
60 0.095
80 0.15
100 0.19
120 0.36
140 0.57
160 0.89
};
\addlegendentry{$\kappa = 10$ SNR =-12}

\addplot [styleDVBTBM, dashdotted]
table {%
40 0.011
60 0.011
80 0.0175
100 0.02
120 0.023
140 0.04
160 0.13
};
\addlegendentry{$\kappa  = 100$ SNR =-12}

\addplot [stylekappa1000, dashdotted]
table {%
40 0.011
60 0.011
80 0.018
100 0.021
120 0.026
140 0.055
160 0.18
};
\addlegendentry{$\kappa  = 1000$ SNR =-12}
\legend{}
\end{axis}

\begin{axis}[%
    width=\fwidth,
    height=\fheight,
    at={(0\fwidth,0\fheight)},
    scale only axis,
    xmin=1,
    xmax=100,
    xtick={},
    ytick={},
    xticklabels={{}, {}, {},{}},
    yticklabels={},
    xtick style = {draw=none},
    ytick style = {draw=none},
    ymin=0.0001,
    ymax= 1,
    legend style={
        /tikz/every even column/.append style={column sep=0.2cm},
        at={(0.18, 0.59)}, 
        anchor=south, 
        draw=white!80!black, 
        font=\scriptsize,
        fill opacity=0.8
        }
]

\addplot [stylekappa10, only marks]
table[row sep=crcr] {%
1	-5 \\
};
\addlegendentry{$\kappa = 10$}

\addplot [styleDVBTBM, only marks]
table[row sep=crcr] {%
1	-5 \\
};
\addlegendentry{$\kappa = 100$}

\addplot [stylekappa1000, only marks]
table[row sep=crcr] {%
1	-5 \\
};
\addlegendentry{$\kappa = 1000$}

\addplot [color=black, dashed, very thick]
  table[row sep=crcr]{%
1	-5\\
};
\addlegendentry{${\rm SNR}=-14$}

\addplot [color=black, dashdotted, very thick]
  table[row sep=crcr]{%
1	-5\\
};
\addlegendentry{${\rm SNR}=-12$}

\end{axis}

\end{tikzpicture}

%% file: figs/pmdtd_vs_kappa.tex
\definecolor{colorDVBTBM}{RGB}{0, 0, 255}
\definecolor{colorkappa10}{RGB}{31, 119, 180}
\definecolor{colorkappa1000}{RGB}{148, 103, 189}  

\pgfplotsset{
    every tick label/.append style={font=\footnotesize},
    every axis plot/.append style={
        line width=1pt, 
        mark size=2,        
        mark options={solid}
    }
}

\tikzset{
    styleDVBTBM/.style={color=colorDVBTBM, mark=*},
    stylekappa10/.style={color=colorkappa10, mark=o},
    stylekappa1000/.style={color=colorkappa1000, mark=otimes},
}

\pgfplotsset{every tick label/.append style={font=\footnotesize}}
\begin{tikzpicture}

\begin{axis}[
    width=\fwidth,
    height=\fheight,
    at={(0\fwidth,0\fheight)},
    scale only axis,
    ylabel style={font=\normalsize},
    xlabel style={font=\normalsize},
    xlabel={$\kappa$},
    ylabel={$p_{\rm md}^{\rm TD}$},
    axis background/.style={fill=white},
    legend style={
    /tikz/every even column/.append style={column sep=0.2cm},
    at={(0.22, 0.02)}, 
    anchor=south, 
    draw=white!80!black, 
    font=\scriptsize,
    fill opacity=0.8
    },
    xmajorgrids,
    xminorgrids,
    ymajorgrids,
    yminorgrids,
    xmin=10, 
    xmax=100000,
    ymin=0.0005, 
    ymax=1,
    ymode=log,
    xmode=log,
]

\addplot [styleDVBTBM]
table {%
10 0.99
100 0.003
1000 0.0028
10000 0.0026
100000 0.00268749
};
\addlegendentry{${\rm SNR} = -5$ dB}
\addplot [styleDVBTBM, dotted]
table {%
10 0.99
100 0.025
1000 0.027
10000 0.031
100000 0.05
};
\addlegendentry{${\rm SNR} = -10$ dB}
\addplot [styleDVBTBM, dashdotted]
table {%
10 0.99
100 0.13
1000 0.18
10000 0.24
100000 0.36
};
\addlegendentry{${\rm SNR} = -12$ dB}
\addplot [styleDVBTBM, dashed]
table {%
10 0.99
100 0.44
1000 0.6
10000 0.74
100000 0.85
};
\addlegendentry{${\rm SNR} = -14$ dB}
\end{axis}

\end{tikzpicture}

%% file: figs/pmdtd_vs_SNR_z.tex
\definecolor{colorDVBTBM}{RGB}{0, 0, 255}
\definecolor{colorz2}{RGB}{227,119,194}
\definecolor{colorz4}{RGB}{0,102,51}
\definecolor{colorz5}{RGB}{101,67,33}

\pgfplotsset{
    every tick label/.append style={font=\footnotesize},
    every axis plot/.append style={
        line width=1pt, 
        mark size=2,        
        mark options={solid}
    }
}

\tikzset{
    styleDVBTBM/.style={color=colorDVBTBM, mark=*},
    stylez2/.style={color=colorz2, mark=*},
    stylez4/.style={color=colorz4, mark=*},
    stylez5/.style={color=colorz5, mark=*}
}
\pgfplotsset{every tick label/.append style={font=\footnotesize}}
\begin{tikzpicture}
\begin{axis}[
    width=\fwidth,
    height=\fheight,
    at={(0\fwidth,0\fheight)},
    scale only axis,
    ylabel style={font=\normalsize},
    xlabel style={font=\normalsize},
    xlabel={SNR [dB]},
    ylabel={$p_{\rm md}^{\rm TD}$},
    axis background/.style={fill=white},
    legend style={
        /tikz/every even column/.append style={column sep=0.2cm},
        at={(0.15, 0.05)}, 
        anchor=south, 
        draw=white!80!black, 
        font=\scriptsize,
        fill opacity=0.8
        },
    xmajorgrids,
    xminorgrids,
    ymajorgrids,
    yminorgrids,
    xmin=-14,
    xmax=-5,
    ymin=0.0035,
    ymax=0.7,
    ymode=log,
]

\addplot [stylez2]
table {%
-14 0.44
-12 0.16
-10 0.08
-8  0.0368
-6  0.0158
-5  0.01
};
\addlegendentry{$z=2$}
\addplot [styleDVBTBM]
table {%
-14 0.44
-12 0.11
-10 0.032
-8  0.0121
-6  0.0054
-5  0.004
};
\addlegendentry{$z=3$}
\addplot [stylez4]
table {%
-14 0.61
-12 0.22
-10 0.045
-8  0.0153
-6  0.0069
-5  0.0055

};
\addlegendentry{$z=4$}
\addplot [stylez5]
table {%
-14 0.67
-12 0.23
-10 0.05
-8  0.0165
-6  0.0069
-5  0.0052
};
\addlegendentry{$z=5$}
\end{axis}

\end{tikzpicture}

%% file: figs/pmdtd_vs_Ka_comp.tex
\definecolor{colorDVBTBM}{RGB}{0, 0, 255}
\definecolor{colorHybALS}{RGB}{128, 0, 128}
\definecolor{colorALS}{RGB}{140, 86, 75} 

\pgfplotsset{
    every tick label/.append style={font=\footnotesize},
    every axis plot/.append style={
        line width=1pt, 
        mark size=2,        
        mark options={solid}
    }
}

\tikzset{
    styleDVBTBM/.style={color=colorDVBTBM, mark=*},
    styleHybALS/.style={color=colorHybALS, mark=square*},
    styleALS/.style={color=colorALS, mark=diamond*}
}

\pgfplotsset{every tick label/.append style={font=\footnotesize}}
\begin{tikzpicture}

\begin{axis}[
    width=\fwidth,
    height=\fheight,
    at={(0\fwidth,0\fheight)},
    scale only axis,
    ylabel style={font=\normalsize},
    xlabel style={font=\normalsize},
    xlabel={$K_{\rm a}$},
    ylabel={$p_{\rm md}^{\rm TD}$},
    axis background/.style={fill=white},
    legend style={
      at={(0.03,0.97)},
      anchor=north west,
      fill opacity=0.7,
      text opacity=1,
      legend cell align=left, 
      align=left,
      font=\footnotesize,
      draw=white!80!black,        
    },
    xtick = {40, 60, ..., 240},
    xmajorgrids,
    xminorgrids,
    ymajorgrids,
    yminorgrids,
    xmin=40,
    xmax=240,
    ymin=0.008, 
    ymax=1,
    ymode=log,
]

\addplot [styleALS, dotted]
table {%
40 0.01
60 0.0122
80 0.014
100 0.05
120 0.97
140 0.99
160 0.99
180 0.99
200 0.99
220 0.99
240 0.99
};
\addlegendentry{ALS SNR=-10}

\addplot [styleDVBTBM, dotted]
table {%
40 0.01
60 0.01
80 0.010
100 0.0101
120 0.011
140 0.0116
160 0.03
180 0.06
200 0.26
220 0.55
240 0.75
};

\addlegendentry{DVB-ALS SNR=-10}

\addplot [styleHybALS, dotted]
table {%
40 0.01
60 0.01
80 0.011
100 0.011
120 0.011
140 0.04
160 0.5
180 0.96
200 0.99
220 0.99
240 0.99
};
\addlegendentry{Hybrid-ALS SNR=-10}

\addplot [styleDVBTBM, dashed]
table {%
40 0.0219
60 0.0201
80 0.035
100 0.052
120 0.08
140 0.18
160 0.44
180 0.56
200 0.75
220 0.87
240 0.88
};

\addlegendentry{DVB-ALS SNR=-14}

\addplot [styleALS, dashed]
table {%
40 0.035
60 0.04
80 0.067
100 0.8
120 0.97
140 0.99
160 0.99
180 0.99
200 0.99
220 0.99
240 0.99
};
\addlegendentry{ALS SNR=-14}

\addplot [styleHybALS, dashed]
table {%
40 0.06
60 0.062
80 0.077
100 0.11
120 0.17
140 0.76
160 0.956
180 0.96
200 0.99
220 0.99
240 0.99
};
\addlegendentry{Hybrid-ALS SNR=-14}

\legend{}
\end{axis}

\begin{axis}[%
    width=\fwidth,
    height=\fheight,
    at={(0\fwidth,0\fheight)},
    scale only axis,
    xmin=1,
    xmax=100,
    xtick={},
    ytick={},
    xticklabels={{}, {}, {},{}},
    yticklabels={},
    xtick style = {draw=none},
    ytick style = {draw=none},
    ymin=0.0001,
    ymax= 1,
    legend style={
        /tikz/every even column/.append style={column sep=0.2cm},
        at={(0.81, 0.02)}, 
        anchor=south, 
        draw=white!80!black, 
        font=\scriptsize,
        fill opacity=0.8
        }
]

\addplot [styleALS, only marks]
table[row sep=crcr] {%
1	-5 \\
};
\addlegendentry{ALS}

\addplot [styleHybALS, only marks]
table[row sep=crcr] {%
1	-5 \\
};
\addlegendentry{Hybrid-ALS}

\addplot [styleDVBTBM, only marks]
table[row sep=crcr] {%
1	-5 \\
};
\addlegendentry{DVB-ALS}

\addplot [color=black, dashed, very thick]
  table[row sep=crcr]{%
1	-5\\
};
\addlegendentry{${\rm SNR}=-14$}

\addplot [color=black, dotted, very thick]
  table[row sep=crcr]{%
1	-5\\
};
\addlegendentry{${\rm SNR}=-10$}

\end{axis}

\end{tikzpicture}

%% file: figs/ura_compa_N=16.tex
\definecolor{colorFASURA}{RGB}{255, 0, 0}
\definecolor{colorDVBTBM}{RGB}{0, 0, 255}
\definecolor{color4PTURA}{RGB}{255, 127, 14}
\definecolor{color3PTURA}{RGB}{0, 128, 0}
\definecolor{colorHybALS}{RGB}{128, 0, 128}

\pgfplotsset{
    every tick label/.append style={font=\footnotesize},
    every axis plot/.append style={
        line width=1pt, 
        mark size=2,        
        mark options={solid}
    }
}

\tikzset{
    styleFASURA/.style={color=colorFASURA, mark=x},
    styleDVBTBM/.style={color=colorDVBTBM, mark=*},
    style3PTURA/.style={color=color3PTURA, mark=triangle*},
    style4PTURA/.style={color=color4PTURA, mark=triangle},
    styleHybALS/.style={color=colorHybALS, mark=square*}
}

\pgfplotsset{every tick label/.append style={font=\footnotesize}}
\begin{tikzpicture}

\begin{axis}[
    width=\fwidth,
    height=\fheight,
    at={(0\fwidth,0\fheight)},
    scale only axis,
    ylabel style={font=\normalsize},
    xlabel style={font=\normalsize},
    xlabel={$K_{\rm a}$},
    ylabel={$E_{\rm b}/N_0$ [dB]},
    axis background/.style={fill=white},
    legend style={
        /tikz/every even column/.append style={column sep=0.2cm},
        at={(0.02,0.97)},
        anchor=north west,
        draw=white!80!black, 
        font=\scriptsize,
        fill opacity=0.8
        },
    ytick = {-6,-3,...,10},
    xtick = {80,100, ..., 240},
    xmajorgrids,
    xminorgrids,
    ymajorgrids,
    yminorgrids,
    xmin=80, 
    xmax=240,
    ymin=-7, 
    ymax=10
]

\addplot [styleFASURA]
table {%
80  -6.13
100 -5.63
120 -5.38
140 -3.78
160 -2.38
180 -0.13
200 8.12
220 24.62
240 24.62
};
\addlegendentry{FASURA}

\addplot [style3PTURA]
table {%
80  -3.38
100 -2.88
120 -0.16
140 7.62
160 25.62
180 24.62
200 24.62
220 24.62
240 24.62
};
\addlegendentry{3-PTURA}

\addplot [styleHybALS]
table {%
80  -2.63
100 -2.38
120 -2.18
140 -1.78
160 -1.13
180 0.37
200 24.62
220 24.62
240 24.62
};
\addlegendentry{Hybrid-\ac{tbm}}

\addplot [styleDVBTBM]
table {%
80  -3.08
100 -2.78
120 -2.58
140 -2.18
160 -1.78
180 -1.13
200 -0.38
220 1.245
240 4.62
};
\addlegendentry{\ac{dvbtbm}} 

\end{axis}

\end{tikzpicture}

%% file: figs/ura_compa_N=50.tex
\definecolor{colorFASURA}{RGB}{255, 0, 0}
\definecolor{colorDVBTBM}{RGB}{0, 0, 255}
\definecolor{color4PTURA}{RGB}{255, 127, 14}
\definecolor{color3PTURA}{RGB}{0, 128, 0}
\definecolor{colorHybALS}{RGB}{128, 0, 128}

\pgfplotsset{
    every tick label/.append style={font=\footnotesize},
    every axis plot/.append style={
        line width=1pt, 
        mark size=2,        
        mark options={solid}
    }
}

\tikzset{
    styleFASURA/.style={color=colorFASURA, mark=x},
    styleDVBTBM/.style={color=colorDVBTBM, mark=*},
    style3PTURA/.style={color=color3PTURA, mark=triangle*},
    style4PTURA/.style={color=color4PTURA, mark=triangle},
    styleHybALS/.style={color=colorHybALS, mark=square*}
}
\begin{tikzpicture}

\begin{axis}[
    width=\fwidth,
    height=\fheight,
    at={(0\fwidth,0\fheight)},
    scale only axis,
    ylabel style={font=\normalsize},
    xlabel style={font=\normalsize},
    xlabel={$K_{\rm a}$},
    ylabel={$E_{\rm b}/N_0$ [dB]},
    axis background/.style={fill=white},
    legend style={
        /tikz/every even column/.append style={column sep=0.2cm},
        at={(0.03,0.97)},
        anchor=north west,
        draw=white!80!black, 
        font=\scriptsize,
        fill opacity=0.8
        },
    ytick = {-12,-10,...,0},
    xmajorgrids,
    xminorgrids,
    ymajorgrids,
    yminorgrids,
    xmin=400, 
    xmax=1300,
    ymin=-12,
    ymax=-3.5
]

\addplot [styleFASURA] table {%
400 -11.37
500 -10.82
600 -10.2
700 -9.5
800 -8.51
900 -7
1000 -5
1100 15
1200 20
};
\addlegendentry{FASURA}

\addplot [style3PTURA] table {%
400 -10.9
500 -10.8
600 -10.6
700 -10.4
800 -9.6
900 -8.3
1000 -6.7
1100 -3.5
1200 15
};
\addlegendentry{3-PTURA}

\addplot [style4PTURA] table {%
400 -9
500 -8.8
600 -8.7
700 -8.6
800 -8.4
900 -7.8
1000 -7.2
1100 -6.2
1200 -3.5
};
\addlegendentry{4-PTURA}

\addplot [styleDVBTBM] table {%
400 -10.0735002168009
500 -9.74850021680094
600 -9.44850021680094
700 -9.14850021680094
800 -8.64850021680094
900 -8.24850021680094
1000 -7.44850021680094
1100 -7.07350021680094
1200 -6.44850021680094
1300 -3.94850021680094
};
\addlegendentry{\ac{dvbtbm}}

\end{axis}
\end{tikzpicture}